\documentclass[10pt,acmsmall,letterpaper,noacm=true,authorversion=true]{acmart}

\usepackage[english]{babel}
\usepackage{blindtext}
\usepackage{multirow}
\usepackage{balance}
\usepackage{placeins}

\setcopyright{none}
\renewcommand\footnotetextcopyrightpermission[1]{} % removes footnote with conference information in first column
\usepackage[format=plain,labelfont=bf,textfont=it,skip=0pt]{caption}

\graphicspath{{figs/}}

\acmMonth{6}
\acmYear{2022}
\acmVolume{1}
\copyrightyear{2021}
\setcopyright{acmcopyright}
\acmConference{SIGMETRICS '22}{Summer Submission}{Mumbai, India.}
\acmPrice{TBA}
\acmDOI{TBA}
\acmISBN{TBA}

\begin{document}
\title{GPS-Based Geolocation of Consumer IP Addresses}

% \if 0
\author{James Saxon, Nick Feamster}
\orcid{0000-0002-5151-4517}
\affiliation{%
  \institution{Department of Computer Science and Center for Data and Computing, University of Chicago}
  \streetaddress{5730 South Ellis Ave}
  \city{Chicago} 
  \state{IL} 
  \postcode{60637}
}
\email{{jsaxon, feamster}@uchicago.edu}
% \fi

\newcommand{\textreg}{\textsuperscript{\textregistered}}
\newcommand{\note}[1]{\textcolor{red}{[#1]}}
\newcommand{\fnote}[1]{\footnote{\textbf{\textcolor{red}{#1}}}}
\newcommand{\missing}{\textbf{\textcolor{red}{[MISSING!]}}}
\newcommand{\linklogo}{\includegraphics[height=0.8em]{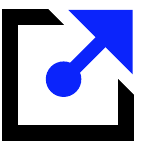}}

\renewcommand{\paragraph}[1]{\vspace{0.35em}\noindent\textbf{#1}}

\hypersetup{
  colorlinks=true,linkcolor=blue,citecolor=blue,urlcolor=blue,
  pdftitle={Heterogeneity in the Localizability of IP Addresses}, pdfauthor={James Saxon}
  pdfpagemode={UseOutlines},
  bookmarksopen=true, bookmarksnumbered=true,
  pdfstartview={Fit}
}

\begin{sloppypar}
\begin{abstract}
This paper uses two commercial datasets
  of IP addresses from smartphones, geolocated through the Global Positioning
    System (GPS),
    to characterize the geography of IP address subnets from 
    mobile and broadband ISPs.
Datasets that geolocate IP addresses based on GPS offer superlative
    accuracy and precision for IP geolocation and thus provide an unprecedented opportunity
    to understand both the accuracy of existing geolocation databases as well
    as other properties of IP addresses, such as mobility and churn.
We focus our analysis on three large cities in the United States.

After evaluating the accuracy
  of existing geolocation databases, 
  we analyze the circumstances under which IP geolocation databases may be
    more or less accurate.
Within our sample, we find that geolocation databases are 
  more accurate on fixed-line than mobile networks,
  that IP addresses on university networks can be more accurately located
   than those from consumer or business networks, and that often the paid
    versions of these databases are not significantly more accurate than the
    free versions.
We then characterize how quickly
    subnets associated with fixed-line networks change geographic locations,
  and how long residential broadband ISP subscribers retain
    individual IP addresses.
We find, generally, that most IP address assignments are stable
  over two months, although stability does vary across ISPs.
Finally, we evaluate the suitability of
  existing IP geolocation databases for understanding Internet 
  access and performance in human populations within specific geographies and
    demographics.
Although the median accuracy of IP geolocation
  is better than 3~km in some contexts~-- 
  fixed-line connections in New York City, for instance~-- 
  we conclude that relying on IP geolocation databases to understand
    Internet access in densely populated regions such as cities is premature.
\end{abstract}

\maketitle

\section{Introduction}

IP geolocation is a longstanding problem in computer networking,
  with both an active academic research and a wide array of commercial
    solutions and applications.
IP geolocation is used for a variety of purposes, including mapping clients to
    nearby content delivery network (CDN) replicas, personalization of search
    results and advertising, and customization of content (e.g., weather,
    language localization).
    In a legal context, IP geolocation is used for digital rights management
    (e.g, geographic licensing restrictions), compliance with
    the laws and regulations of a region or country (e.g., gambling, sales
    taxes, privacy regulations), and to assist with law enforcement
    (e.g., determining jurisdictions or collecting evidence).
In security contexts, government and commercial entities use it for
  for counter-terrorism, attack attribution,
  monitoring access to private networks, and detecting potential fraud.
It facilitates operations and site reliability (e.g., monitoring packet loss from a location), and
  informs infrastructure investments by both industry and
    policymakers~\cite{2010_gill_geolocation_circumvention,2006_gueye_constraint_based_geolocation,2018_ovidiu_geolocation_reverse_dns,2011_poese_uhlig_ip_geolocation_unreliable,2013_li_geolocation_moderately_connected,2011_wang_street_level_geolocation}.
Computer science researchers also use IP geolocation to study the properties
and evolution of the network itself, such as the structure and graph parameters of networks~\cite{2005_freedman_feamster_geographic_locality_ip_prefixes,2002_spring_rocketfuel_geo}.

Increasingly, IP geolocation is being used to address various problems in
{\em policy and social science} that entail drawing inferences about various
demographics and geographies based on inferred locations of IP addresses.
Social scientists have noted the potential to use ``big data'' as a lens on
human behaviors and
interactions~\cite{2009_lazer_computational_social_science}, and as 
  modern society is increasingly mediated through the
  Internet, many of our interactions are associated with IP addresses.
Server logs and speed measurements, for instance,
  show who accesses resources and the quality of their connections.
This allows aggregate statistics or time trends.
But associating these behaviors and network conditions with human populations
  ultimately requires a way to map IP addresses to physical locations. 
A natural approach would be to use IP geolocation 
  with census tract-scale precision to link IP addresses to physical locations.
In this paper, we evaluate whether 
  free and paid IP geolocation databases
  can achieve this level of accuracy, in large cities in the United States.
We also explore the determinants of IP geolocation accuracy.

The accuracy of IP geolocation databases has practical implications for the
answers to a wide range of social and public policy questions.
One area of particular timeliness is that of the so-called ``digital divide.''
Calls for digital equity and inclusion, already urgent,
  have reached a fever pitch during the COVID-19 pandemic.  
Prominent studies of broadband performance
  from Microsoft and M-Lab rely on IP geolocation
  to associate Internet throughput and latencies with zip codes~\cite{2021_microsoft_broadband,2021_mlab_visualization}.
\citeauthor{2019_ganelin_chuang_ip_mooc_regressive} studied whether or not
geolocation databases could reliably indicate socioeconomic status of MOOC
registrants with known physical addresses.
That study ultimately concluded,
as will we, that answering such questions based on existing IP geolocation
databases is premature~\cite{2019_ganelin_chuang_ip_mooc_regressive}.

We revisit this problem now,
  due both to its practical implications,
  and thanks to the availability of two highly-accurate and large-scale
  groundtruth datasets of GPS-located IP addresses.
These datasets, from Unacast and Ookla\textreg{} Speedtest Intelligence\textreg{},
  afford us a view of consumer behaviors on both fixed-line and mobile networks,
  that is markedly different from the geolocation targets used in past work.

Table~\ref{tab:findings} presents our main findings.
The rest of this paper is organized as follows.
Section~\ref{sec:related} discusses related work in IP geolocation, both in
research and in commercial product offerings.
Section~\ref{sec:data} describes 
  the datasets that we use for the analysis in this paper.
Section~\ref{sec:quality} evaluates
  the quality of the datasets that we are using,
  in particular exploring the suitability of using GPS data as a ``ground truth''
  for evaluating IP geolocation databases.
Section~\ref{sec:results} presents the result of our study,
  including findings about the circumstances under which IP geolocation is more or less accurate.
In Section~\ref{sec:discussion},
  we interpret and extend our results in the context of
  research on human populations and privacy. We conclude in
  Section~\ref{sec:conclusion}.

\begin{table}
\centering
\fontsize{9}{11} \selectfont
\setlength{\tabcolsep}{2pt}
\begin{tabular}{cp{0.9\linewidth}}
\hline
\S & Main Findings \\
\hline
\ref{ssec:groundtruth} & GPS reports are a credible groundtruth of IP locations. \\
\ref{ssec:db_error}    & MaxMind's GeoIP2 service provides the lowest median error among tested services and cities: 2.62~km on fixed-line addresses in NYC.\\
\ref{ssec:reliable}    & IP geolocation performs better on fixed-line consumer networks and universities, and worse on mobile broadband and businesses. \\
\ref{ssec:size}        & The physical size of subnets is correlated with the accuracy with which they are IP geolocated. \\
\ref{ssec:movement}    & On the two-month time-scale, the median fixed-line /24 IPv4 subnet in US cities moves less than 1~km. \\
\ref{ssec:churn}       & Churn of individual IP addresses on fixed-line networks in major US cities takes months. \\
\ref{sec:discussion}   & Access modalities -- mobiled vs fixed -- differ between demographic groups.  Even on fixed-line networks, relying on IP geolocation to identify neighborhoods would lead to biased results. \\
\hline
\end{tabular}
\caption{Main findings, with pointers to sections. \label{tab:findings}}
\end{table}

\section{Related Work}\label{sec:related}

Past work on IP geolocation generally takes three approaches,
as outlined by
\citeauthor{2001_padmanabhan_geoping_geocluster}~\cite{2001_padmanabhan_geoping_geocluster}.
Their IP Geolocation work, IP2Geo, compared the complementary strengths of
active latency measurements (GeoPing),
active traceroutes paired with DNS hints (GeoTrack), and
static databases of outside information (GeoCluster).
Each of these approaches has evolved.
Padmanabhan and Subramanian
concluded that database-driven methods held the greatest promise.
Commercial products have accordingly built
databases with proprietary methods that
include registry information, outside data, and active methods.
On the other hand, academic work has tended to focus
on active and DNS-based measurements.

\paragraph{IP geolocation methods.} Starting with DNS,
\citeauthor{2002_spring_rocketfuel_geo} developed techniques in their
Rocketfuel project to map infrastructure (i.e., routers) to physical locations.
A significant contribution was to optimize traceroute targets to
  minimize redundancy and ensure that each path will traverse its target
  ISP~\cite{2002_spring_rocketfuel_geo}, although their use of the DNS to
  geolocate routers was pioneering at the time.
Their subsequent approach to DNS hint identification was largely manual~--
  ``browsing through the list of router names''~--
  but the resultant \texttt{undns} tool has proven influential and
enduring.  \citeauthor{2005_freedman_feamster_geographic_locality_ip_prefixes}
extended \texttt{undns}'
coverage~\cite{2005_freedman_feamster_geographic_locality_ip_prefixes}.  These
projects were driven by questions about properties of the network,
specifically the topology of large ISPs and the efficiency of block
assignments in BGP routing tables.  More recently,
\citeauthor{2018_ovidiu_geolocation_reverse_dns}~\cite{2018_ovidiu_geolocation_reverse_dns}
attempted to enumerate all possible DNS city name hints and finalize location
decisions with machine learning.
Like IP2Geo, the authors relied on a large dataset
  from Microsoft for their ground truth, although the ground truth data
  was from Bing instead of Hotmail.

In the latency-based space,
\citeauthor{2006_gueye_constraint_based_geolocation}~\cite{2006_gueye_constraint_based_geolocation} and
\citeauthor{2006_katz-basett_geolocation_delay_and_topology} \cite{2006_katz-basett_geolocation_delay_and_topology}
introduced constraint-based geolocation (CBG) and topology-based geolocation (TBG).
CBG is essentially the intersection of several latency-derived distance buffers,
while TPG also localizes intermediate hosts
so that targets can be constrained by their relation to passive landmarks rather than just active probes.
Subsequently, Octant 
  incorporated both positive \emph{and negative} constraints
  (the IP address is \emph{not} within a certain
  radius)~\cite{2007_wong_octang_geolocation_negative}.

In addition to this ``geometric'' approach
are several statistical strategies.
\citeauthor{2010_erikkson_barford_learning_based_geolocation} developed
  first a Bayesian approach and
  then a likelihood-driven choice 
  among possibilities with the CBG-derived
  regions~\cite{2010_erikkson_barford_learning_based_geolocation,2012_eriksson_barford_posit_lightweight_geolocation}.
Other work presents strategies using kernel density and maximum likelihood
estimation~\cite{2009_youn_kde_geolocation,2010_arif_mle_geolocation}.
It is also possible to constrain location
from the covariance matrix of latency measurements with locations.

Notable in \citeauthor{2010_erikkson_barford_learning_based_geolocation}'s Bayesian work
is the insight that outside information can help constrain or inform geolocation.
They used population as a measure of places' importance,
as have later researchers~\cite{2018_ovidiu_geolocation_reverse_dns}.
Other forms of information help as well.
In trace-based work reminiscent of TPG,
  \citeauthor{2011_wang_street_level_geolocation} performed extensive
  webscraping and analysis to identify and confirm businesses with locally-hosted sites
  that they could ``enlist'' as passive landmarks.
They used those landmarks to identify the locations of routers near the
geolocation target~\cite{2011_wang_street_level_geolocation}.

Scalability has long been a limitation of active measurements.
Since locations are most-constrained by the closest locations,
\citeauthor{2012_zi_geolocations_of_millions_of_addresses} developed methods
to prioritize measurements from nearby hosts, effectively by localizing
avatars from subnets~\cite{2012_zi_geolocations_of_millions_of_addresses}.
Alternatively, \citeauthor{2013_li_geolocation_moderately_connected}
  ``flip'' the standard infrastructure of active geolocation with GeoGet:
  the targets to be localized measure the latency themselves, through javascript,
  rather than generating pings through an
  API~\cite{2013_li_geolocation_moderately_connected}.
This reduces the number of servers and traffic required,
  and it is also helpful since clients' devices or networks
  may fail to respond to pings or complete traceroutes.

\paragraph{Evaluating commercial services.}
These advances notwithstanding,
commercial geolocation tends to be implemented through databases, which are
inexpensive to distribute and
can aggregate historical observations across many sources.
The leading services---MaxMind, IP2Location, Akamai, or NetAcuity---all use proprietary methods.
A number of papers assess the performance of these databases,
  comparing with the preceding active methods \cite{2007_imprecision_of_block_geolocation},
  points-of-presence paired with routing tables from a large ISP \cite{2011_poese_uhlig_ip_geolocation_unreliable},
  DNS lookups paired with ground truth rules from domain operators \cite{2017_gharaibeh_router_geolocation},
  from RIPE ATLAS built-in measurements, or PlanetLab nodes,
  or against each other, sometimes with a majority logic applied.
The databases are themselves
  often taken as the ground truth for latency-based measurements
  often with a sort of majority
  logic.
  \citeauthor{2011_shavitt_geolocation_study}
  employ that strategy in evaluating the databases themselves,
  but also focus on \emph{consistency}
  among addresses determined to share a point-of-presence,
  based on an earlier algorithm~\cite{2011_shavitt_geolocation_study,2012_feldman_pop_geolocation}.
Similarly, \citeauthor{2011_huffaker_geocompare}
  assess the agreement of country determinations and distances from a centroid,
  from majority votes (supplemented by PlanetLab ground-truth and limited
  round-trip time measurements)~\cite{2011_huffaker_geocompare}.

On the whole, both the formal literature and ``popular wisdom"
paint a fairly pessimistic picture of geolocation performance.
Research studies from about ten years ago assessed median accuracy of these services 
at 25~km in Western Europe and 100~km in the United States.
On the commercial side,
  \citeauthor{2011_poese_uhlig_ip_geolocation_unreliable}
  quote median accuracies
  between tens and hundreds of kilometers
  for MaxMind and
  IP2Location~\cite{2011_poese_uhlig_ip_geolocation_unreliable}.
Other early works present distributions with ranges between
  hundreds or thousands of kilometers~\cite{2011_shavitt_geolocation_study}.
\citeauthor{2017_gharaibeh_router_geolocation}
  present results for routers in particular,
  with median accuracies between 10~km for NetAcuity and 1,000~km for IP2Location,
  on either extreme of the free and paid versions of MaxMind.
More recently, \citeauthor{2018_ovidiu_geolocation_reverse_dns}
  presented medians between 10 and 30~km, depending
  on the sample and service. \cite{2018_ovidiu_geolocation_reverse_dns}
They present results in 10~km bins and do not differentiate
  performance at the very bottom of the range.

\paragraph{Studies of how Internet infrastructure affects geolocation accuracy.}
A persistent though somewhat more subtle current
of the literature has explored the physical
structure of the Internet and its relation to geolocation accuracy.
\citeauthor{2001_padmanabhan_geoping_geocluster} anticipated
  the interplay between network infrastructure and geolocation accuracy in
  2001~\cite{2001_padmanabhan_geoping_geocluster}.
They noted the impact
  of the geographical concentration of AOL's login nodes on accuracy,
  and showed that clusters of addresses that were physically larger
  were associated with poorer performance for the GeoCluster (database) method.
This point was echoed in 2007
  \citeauthor{2007_imprecision_of_block_geolocation} \cite{2007_imprecision_of_block_geolocation}
Similarly, \citeauthor{2005_freedman_feamster_geographic_locality_ip_prefixes}
  measured the physical scale of autonomous systems.
Later, \citeauthor{2016_gharaibeh_geo_ip_colocality}
  probed the common assumption of databases
  that /24 subnets are co-located~\cite{2007_imprecision_of_block_geolocation}
Those papers show that
  systems, subnets, and IP prefixes advertised by the Border Gateway Protocol
  (BGP)
  can span large physical distances.
In this paper, we seek to extend this work, aiming to identify the circumstances when they are large or small.
\citeauthor{2011_huffaker_geocompare}
  characterized accuracy according to carriers' network role;
  we extend that line of exploration in this research, exploring how accuracy varies
  between commercial ISPs, large companies, and universities.
We categorize addresses by ``Doing-Business As" names reported
in IP address registries; to our knowledge, such a characterization is
unprecedented, at least in the current era where mobile devices are
significantly more prevalent than they were a decade ago.

In addition to work on IP address \emph{locations},
  our data also shed light on the persistence of 
  dynamically assigned IP addresses, itself an active area of analysis.
Recent works have used RIPE Atlas probes~\cite{2016_padmanabhan_reasons_ips_change,2020_komosny_ip_address_survival},
  javascript-based user monitoring by a large CDN \cite{2020_padmanabhan_dynamips_ip_assignment},
  and browser extensions \cite{2020_mishra_ip_address_retention}
  to study address retention times.
Times range from nearly-ephemeral on mobile networks, 
  to many months for fixed-line connections in the North America.
The retention times we observe are broadly 
  consistent with previous findings for North America.

Finally, our project is 
  informed by recent work on Carrier Grade Network Address Translation (GC-NAT).
CG-NATs are increasingly common across all ISPs,
  but almost ubiquitous on mobile carriers \cite{2016_richter_carrier_grade_nat}.
Since we geolocate public IP addresses,
  it stands to reason that the geolocation accuracy of devices behind a CG-NAT
  cannot be more precise than the basic spatial scale over
  which a public address is used.
Nevertheless, the physical extent of CG-NATs' structures have not been studied.
Public IP addresses could map to limited geographic locations
  like antennas, or to larger ones like cities.

\paragraph{How this paper extends past work.}
Past work that evaluates IP geolocation accuracy has tended to rely either
  on active measurements of somewhat coarse precision,
  or on a fairly consistent set of (unrepresentative) benchmarks:
    specifically, PlanetLab sites and university clusters.
The dataset we rely on for this paper of course has its own peculiarities---it
is a non-random sample of mobile devices---
  but this view from the access network, including mobile devices, is critical
  and distinctive from past studies.
It is a large sample,
  indicative of realistic consumer geolocation targets
  in major cities in the United States.
The Global Positioning System~(GPS)
  has long served as a counterpoint to IP geolocation,
  both as a benchmark of accuracy and as an analog in multilateration.
Historically,
  its deployment and use for Internet measurement
  felt impossibly far off~\cite{2006_katz-basett_geolocation_delay_and_topology,2010_erikkson_barford_learning_based_geolocation,2012_eriksson_barford_posit_lightweight_geolocation},
  but the future has now arrived.

This paper complements and extends previous work
as a result of its large sample of
consumer smartphone locations on diverse networks.
The primary dataset was provided by Unacast;
  we confirm our basic findings with a smaller, Chicago-only
  sample of GPS-located Speedtest\textreg{} data from Ookla\textreg{}.
Similar datasets are readily available
  for commercial applications and academic research.
We exploit this sample to understand how IP geolocation accuracy varies by
geography, carrier, mode of access, and other factors.
In contrast to previous work, which has
  tended to question the overall reliability of geolocation even at country-level accuracy,
  we find that it works fairly well in predictable and well-defined contexts.
Nevertheless, the imperfect accuracy and context-specific performance
  still currently constrain the applicability of IP geolocation for studying
  Internet access by human populations.

\section{The Data}\label{sec:data}

\begin{figure}
\centering
\includegraphics[width=0.4\linewidth]{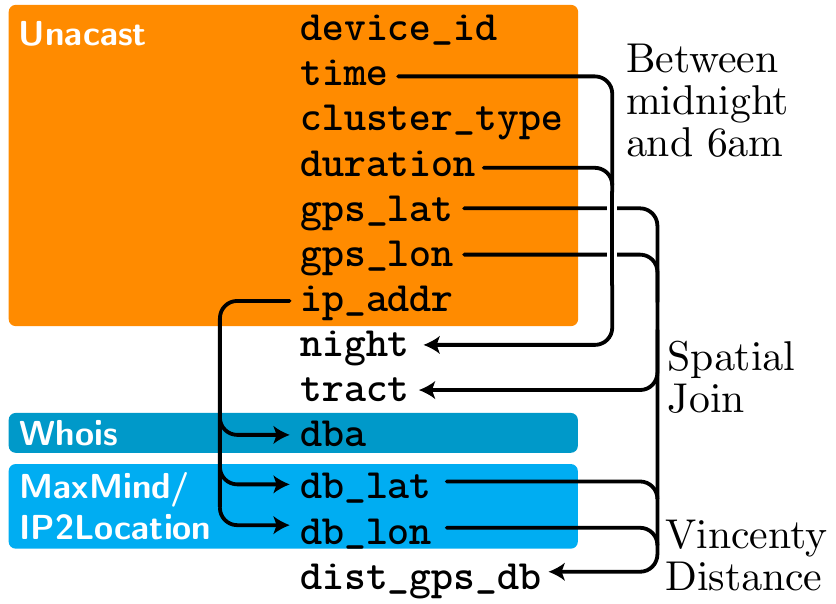}
\caption{Simplified illustration of the data augmentation process, for Unacast data.
  The fundamental data consist of device identifiers, times, locations, and IP addresses.
  Clusters (see text) are also labelled by type, for instance,
    \texttt{TRAVEL} or \texttt{LONG\_AREA\_DWELL}.
  The time and duration are used to construct a flag for night-time clusters.
  The IP address is used with the ARIN whois resource to construct Doing Business As (DBA) names,
    and database-defined locations are retrieved from up to four databases by 
      MaxMind and IP2Location.
  Vincenty distances are calculated 
    between database and GPS locations
}
\label{fig:data}
\end{figure}

This paper relies on two commercial datasets 
  with GPS-tagged IP addresses to analyze
  the geography of consumer IP addresses.
We also evaluate and analyze the performance of 
  databases for IP geolocation from two popular,
  commercial services: IP2Location and MaxMind from the same time periods.
Table~\ref{tab:data} lists the datasets that we use, and
  Figure~\ref{fig:data} illustrates how these datasets are
    joined and augmented in our analysis.

\begin{table}
\centering
\fontsize{9}{11} \selectfont
\begin{tabular}{lccc}
\hline
 & \multirow{2}{*}{Geography} & \multirow{2}{*}{Date} & Location \\
 & & & Reports  \\
\hline
\multirow{2}{*}{Unacast Clusters}  & NYC, Chi., Phl & Aug-Oct 2020 & 248M \\
                        & + 40~mi buffer  & Apr 2021 & 9.5M \\ \hline 
Speedtest Intelligence  & Chicago Region & 2020 & 4M \\ \hline
\multirow{2}{*}{MaxMind Free} & Global & Aug 2020 & - \\
                        & Global & Apr 2021 & - \\
MaxMind Paid            & North America & Apr 2021 & - \\ \hline 
\multirow{2}{*}{IP2Location Free}        & Global & Aug 2020 & - \\
                        & Global & Apr 2021 & - \\
IP2Location Paid        & Global & Apr 2021 & - \\
\hline
\end{tabular}
\caption{Data sources: geographic and temporal coverage, and data volumes (for GPS data only). \label{tab:data}}
\end{table}

The GPS data were delivered anonymized and remain so.
The data were collected in accordance with local laws and opt-out policies~(GDPR),
  and analyzed with approval from our university's Institutional Review Board~(IRB).
The IRB approved analysis of reconstructed ``home locations" for earlier work,
  but emphasized the sensitivity of doing so.
For that reason, we avoided geographic analysis of individual devices in this project,
  and proxied ``residence" simply as activities recorded at night.

\subsection{Unacast GPS Smartphone Locations}
\label{ssec:unacast}

The primary dataset used for the analysis is from Unacast,
  a location intelligence firm.
This dataset contains GPS locations reported by mobile devices,
  along with timestamps and unique, anonymous identifiers.
Unacast aggregates multiple location data streams from other firms;
they perform extensive data validation, de-duplication, and processing
on those streams.
The exact applications that generate locations are not provided.
The share of data reporting IANA
  reserved or private addresses low, 0.5\%,
  and the share of addresses associated with foreign Internet registries
  totals just 0.2\% (mostly RIPE, breakdown shown in the Appendix).
The traffic observed in the Unacast dataset is overwhelmingly IPv4, at 99.6\%.

We use data contained
  within a 40~mile radius of
  three major cities in the United States:
  New York, Chicago, and Philadelphia.
This large buffer includes
  both urban and rural populations.
Two samples were provided:
The first is from August--October 2020.
A second, shorter period from April 2021
  is aligned with licenses for paid geolocation databases to allow us to
  evaluate the accuracy of those services.
As discussed below, the IP address from 
  which a physical location is reported
  is recorded for about half of clusters in the 2020 sample, 
  although this falls to just 15\% in the 2021 sample.
Data are used only when they contain an IP address, so 
  the full dataset thus offers IP addresses
  recorded at over 248 million locations.
Of course, individual IP addresses may be reported many times.

The data also report an estimate of the GPS-based location accuracy;
  the median reported accuracy is 17~meters on the 2020 sample, and 
  11~meters on the 2021 sample.
A small fraction of data (1.7\%) are recorded
  with four or fewer decimal points of coordinate precision,
  corresponding to a physical distance of about 10~m.
We exclude these data from subsequent analyses,
  along with location reports with estimated accuracy worse than 50~m.
We also exclude the small fraction of addresses
  associated with private IP ranges and foreign NICs.

Each line of data represents a \emph{cluster} of 
  location reports, called \emph{bumps}.
Clusters are built by combining bumps
  from an individual device that are close in both time and space,
  using Unacast's proprietary algorithm.
That algorithm uses machine learning to account
  for variation in physical scale among locations:
  a mall is larger than a coffee shop or a home.
Clusters are labelled according to their durations,
  which are also reported.
Locations recorded during movement
  are labelled as \texttt{TRAVEL}.
See the Appendix for a listing of cluster frequencies.
This clustering reduces the data volume
by a factor of 20 while retaining most of the information.
Just as important, Unacast's data licenses with \emph{its} suppliers
  often preclude re-licensing the raw, un-clustered data.

The clustering entails some subtlety:
a single physical location and IP address is reported per cluster,
and thus the centroid of a \texttt{TRAVEL} cluster may not exactly
coincide with the moment that the reported IP address was used.
Indeed, the physical location of a consumer IP address is often not fixed;
for instance, consumers can roam freely through their home while connected to their Wi-Fi.
In practice, individual IP addresses
are recorded at many physical locations---and these locations may be close or distant from each other.

As a means of selecting residential location reports, 
  we flag clusters generated at night.
Night-time clusters are those 
  for which the period between the first and last bumps
  extends into the hours between midnight to 6am of any day.
These clusters represent
  just 4.7\% of clusters but 26\% of bumps.
Only 18\% of devices have at least one night-time cluster,
  but those devices generate the vast majority of the data:
  80\% of clusters and 88\% of bumps.
In short, weighted by data volume,
  most devices have observations at times
  when they can reasonably be assumed to be at home.
For the set of devices with night-time clusters,
  the ratio of devices to the population of the study region is about one
  device for every 20 people.

To investigate the determinants of
  geolocation accuracy, we also identify ISPs.
Each address is associated with its /24 subnet,
  whose organization is retrieved from the ARIN whois registry,
  on September 1 2020, or April 25 2021.
If the prefix size of the associated CIDR block exceeds 24 on IPv4 or 48 on IPv6,
  we follow whois' link to the ``parent'' network.
This strategy is similar in intent to an ASN lookup,
  and we include an ASN-based breakdown of ISPs in the Appendix.
The whois look-up differs in practice primarily in superior coverage of
  the Department of Defense NIC and wireless carriers (AT\&T and T-Mobile),
  especially for the RouteViews databases from August 2020.
Further, the ASN lookup also ``fractures'' organizations
  like small city governments or businesses from their providers.
We associate large and common organizations with standardized ``Doing Business As'' (DBA) names,
  taking particular care to capture the major ISPs in each market (Comcast, Charter, etc.).
We separate AT\&T's and Verizon's mobile broadband from their fixed offerings
  based on the words ``Mobility" or ``Wireless" in the organization name.
This may not be a perfect division:
  ``Verizon Business" and ``AT\&T Services" may include mobile offerings,
  but examining the ASN tables suggests this is not their primary use.
It is worth noting that the sample is dominated
  by locations recorded while connected through mobile providers:
  there are ten times as many locations on AT\&T mobile than AT\&T fixed-line services, 
  and more than five times as many on Verizon mobile than Verizon fixed-line.
However, as we will separate addresses by ISP,
  this sample volume effect is largely ``partitioned out."
Ultimately, each address is associated with a single DBA name for analysis.

These procedures also identify large companies and institutions, in particular, universities.
We flag addresses from universities with at least ten thousand students,
  and Fortune 100 companies.
University clusters are ``classic" targets for academic work on geolocation,
  since they have have meaningful and well-known locations,
  but they are not representative of the consumer space.
We exclude ISPs, including Google, from the Fortune 100 set.
We tabulate IANA special use and non-ARIN addresses,
  as checks on the underlying data,
  but exclude these from subsequent analysis.

\subsection{Geolocated Ookla Speedtest Data}

In addition to the data from Unacast,
  we have obtained Speedtest data from Ookla.
The data are for tests performed on smartphones,
  again with locations from GPS.
This dataset is substantially smaller,
  and is limited in geographic extent to the counties surrounding Chicago.
We appeal to these data as a cross-check of the Unacast
  data that, though more voluminous, were not designed for this work.

We have received over 4 million
  individual Speedtest measurements for 2020,
  though only 270 thousand
  match the period of the study (August 2020).
Unlike Unacast data,
  each location comes from a single moment in time
  (it is not a cluster).
On the other hand, the Speedtest data
  include only the first three bytes of the IP address,
  due to privacy restrictions.
We rely on Ookla's coding of Internet Service Providers.

\subsection{Geolocation Databases and Distances}

We obtain the free versions of the MaxMind and IP2Location databases, for August 1, 2020.
We also acquire both the free and paid versions 
  of these databases from April 26, 2021.
The NetAcuity and Akamai geolocation services,
  which are much more expensive, are not included in this work.
Using these databases,
  we geolocate IP addresses from the GPS sample.
Per the license, this is done only
  for the months of GPS data matching the databases
  (August 2020 and April 2021).

We then measure the Vincenty distance (on the ellipsoid of Earth)
  from each IP-geolocated point to the location recorded by the GPS-enabled device.
For most of what follows, we take the centroids of the GPS clusters
  as the ``ground truth'' and call the entire
  distance the ``accuracy'' or ``error.''
Since the database providers acknowledge their limited resolution
  and in certain cases quantify it accurately,
  this language is perhaps unfair:
  it is different for a database to acknowledge a location as unknown
  or indeterminate (as in reserved, private addresses) than to be ``wrong'' about the location.
Moreover, the GPS data themselves do have 
  some limitations, noted below.
Semantics aside,
  the balance of this work tabulates distances with respect to the ground truth
  and seeks to explain their heterogeneity.

\section{Evaluating Data Quality}\label{sec:quality}

Before coming to questions about the properties of consumer IP addresses, 
  we analyze the quality of our data. We first explore the consistency of the
GPS-based location data we obtain from Unacast and Ookla
  by comparing the data against each other, with respect to geolocation databases.

\subsection{Are GPS data a credible ground truth of IP address locations?}
\label{ssec:groundtruth}

The accuracy of IP geolocation is central to Unacast's core business, and the
company dedicates
enormous resources to validating and maintaining their incoming data streams.
While GPS data from smartphones is generally understood to be accurate,
  datasets from smartphone-based services do often
  incorporate additional data to assist with locating devices
  in circumstances where GPS does not work (e.g., indoors).
Thus, while we expect these GPS-based datasets to be reasonably accurate in general,
  it behooves us to explore the quality of these datasets before proceeding with other questions.
Since we aim to use these datasets as ``ground truth'',
  this analysis may seem a bit of circular.
Our strategy is to compare the \emph{consistency} 
  of IP geolocation result for different
  GPS contexts and across independent GPS samples (Unacast and Ookla).
Of course, this analysis does not exclude the possibility of systematic errors
  arising in \emph{both} GPS datasets, or across all datasets,
  but given the lack of further ground truths, we are left with consistency checks.
Note that direct cross-checks of IP addresses' locations between the two samples are not possible,
  because the Ookla data report subnets not unique addresses,
  and many IP addresses are recorded at multiple physical locations.

\paragraph{Evaluating cluster types.}
The correspondence between GPS coordinates
  and the physical location of its IP address may not be perfect.
For example, we expect that the clustering procedures could affect
  the ``compatibility'' of the IP address and GPS location.
Further, if a GPS location is recorded when no network is available,
  it may be subsequently \emph{reported} at a different physical location
  where an IP address can be obtained.
We would expect these effects 
  to be most severe for \texttt{TRAVEL} clusters,
  as previously discussed.
The flip side of this argument is that 
  navigation applications are more likely to be active
  during \texttt{TRAVEL}.
These apps record location more frequently,
  which could \emph{improve} accuracy.

To evaluate the effects of imperfect knowledge of locations,
  stemming from these effects, we contrast \texttt{TRAVEL} clusters with others.
We will show below that geolocation performance differs by network.
Obviously, it is easier to ``travel" when connected to a mobile than fixed-line network.
We therefore focus this check on a single, mobile network: AT\&T Mobility.
We do observe that accuracy is worse for travel than non-travel data,
  but the difference at the median is only about 2.5\%, for either IP2Location or MaxMind.
As can be seen in the Appendix, the cumulative distribution functions
  for travel and non-travel clusters are fairly across their entire domain.

\begin{figure}
\centering
\includegraphics[width=0.4\linewidth]{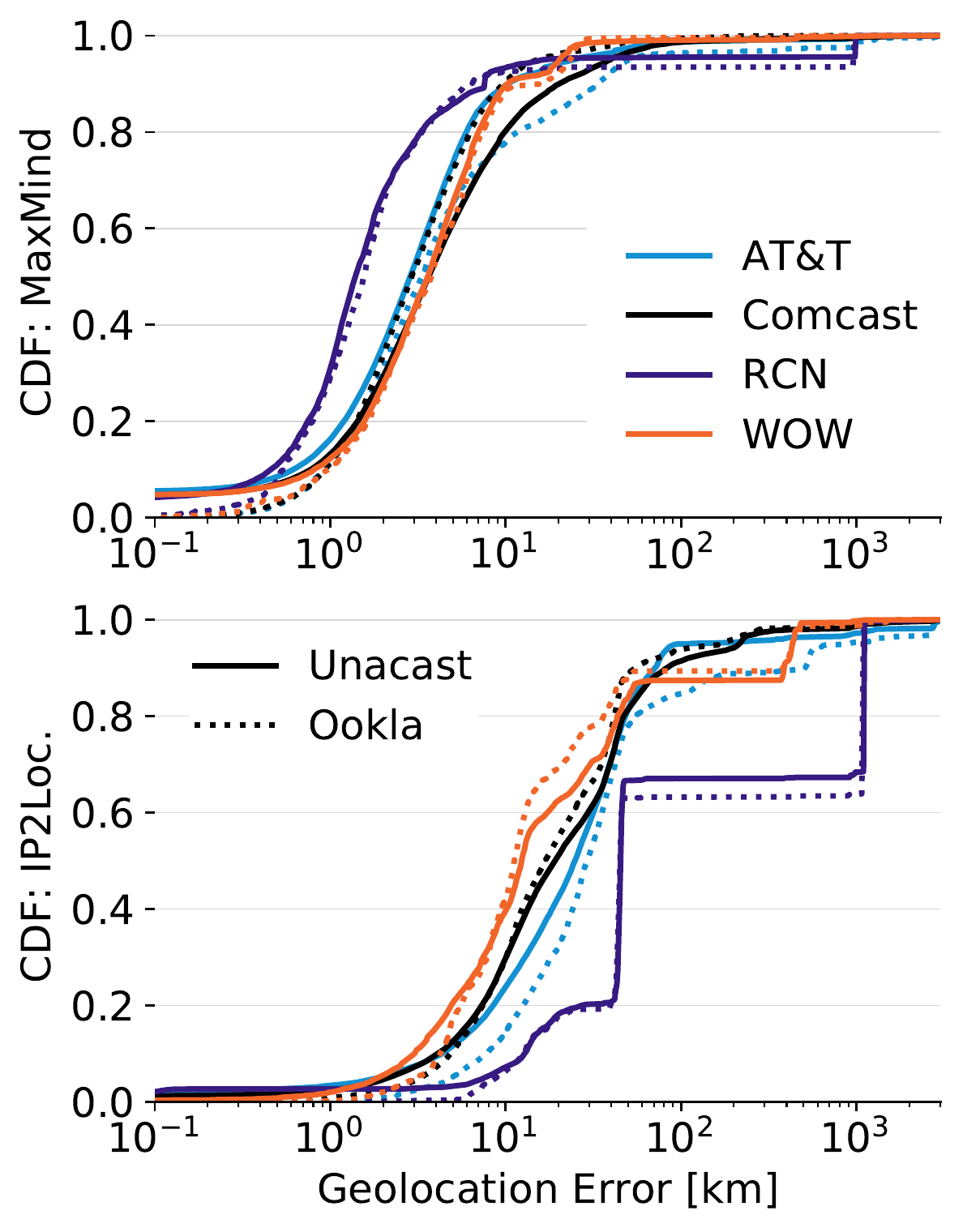}
\caption[Geolocation error of GPS location targets in Chicago, on both Unacast and Ookla Speedtest Intelligence\textreg{} data, using the free versions of the MaxMind and IP2Location databases for August 2020.]{Geolocation error of GPS location targets in Chicago, on both Unacast and Ookla Speedtest Intelligence\textreg{} data,\footnotemark{} using the free versions of the MaxMind and IP2Location databases for August 2020.}
\label{fig:ookla}
\end{figure}

\paragraph{Analysis of independent samples.}
To further validate the GPS data,
  we contrast data from Unacast with Ookla,
  for fixed-line broadband ISPs, in 
  Chicago and August 2020, where both datasets are available
  and aligned with the free versions of the geolocation databases.
Figure~\ref{fig:ookla} shows these results:
  the CDF of location reports as a function of geolocation accuracy.
MaxMind performs somewhat better on Comcast addresses
  from the Ookla dataset than the Unacast data,
  and somewhat worse on AT\&T;
  RCN and WOW! are very consistent.
Discrepancies are somewhat larger on IP2Location
  as is comparative performance by the two databases.

\footnotetext{Based on the authors' analysis of Ookla\textreg{} Speedtest Intelligence\textreg{} data for August 2020 in Chicago. Ookla trademarks used under license and reprinted with permission.}

One notable feature in the 2020 Unacast dataset
  is a small but non-negligible share of the data 
  with IP geolocation ``error" \emph{very} close to zero.
Depending on the ISP,
  that share is 4-5\% of the fixed-line locations on MaxMind
  and 1-2\% of those on on IP2Location.
On close inspection, these appear to be locations
  reported by applications \emph{relying the IP Geolocation services themselves},
  rather than true GPS coordinates.
For example, these ultra-``accurate" locations 
  are not at residences, as one might expect for fixed-line ISPs,
  but in parks, as is MaxMind's practice for default locations~\cite{2020_mishra_ip_address_retention,2020_komosny_ip_address_survival}.
The share of ``too-close" locations is smaller on the 2021 clusters;
  however, the IP address field is populated for a lower share of those data.

However, the basic features of Figure~\ref{fig:ookla} 
  are consistent in the completely separate sample from Ookla,
  which does not exhibit this feature.

\subsection{Which database provides the lowest error in location?}
\label{ssec:db_error}

The practical question is which database to use,
  and how well it should be expected to perform.
This analysis, uniquely, is performed
  using the April 2021 sample from Unacast,
  for which the paid geolocation databases were licensed.
Since Section~\ref{ssec:reliable} will show that
  geolocation on mobile broadband is very poor,
  this analysis focusses on fixed-line broadband.

\begin{figure}
\centering
\includegraphics[width=0.4\linewidth]{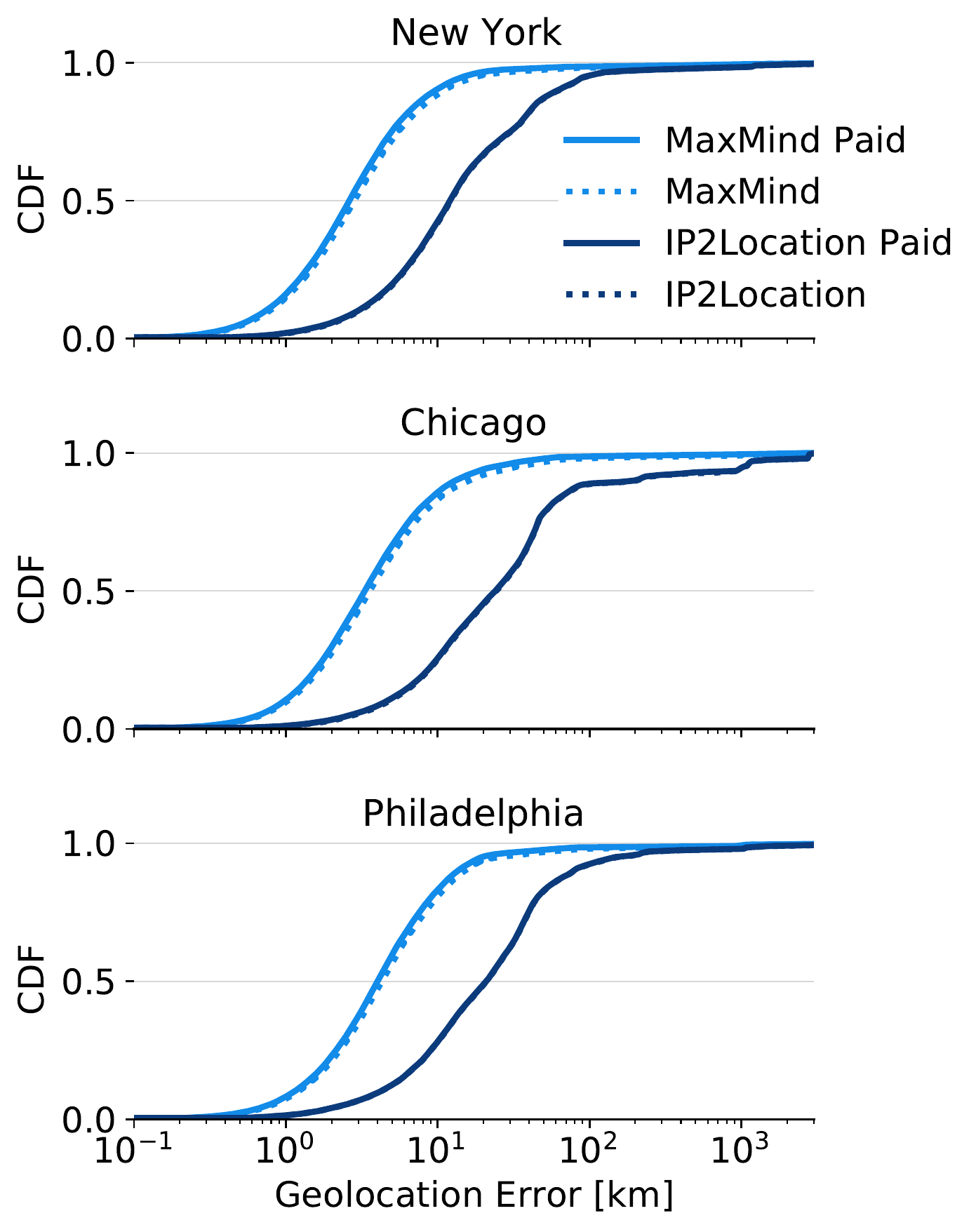}
\caption{Cumulative distribution function by geolocation database and city. Colors reference databases, and line styles denote paid and free versions. \label{fig:cdf_by_city}}
\end{figure}

The short answer is that MaxMind's paid database, GeoIP2, provides the best
accuracy, in terms of geolocation error on all quantiles.
The traditional way of reporting this is the median error,
  which is 2.62~km in New York City, 3.31~km in Chicago, 
  and 4.02~km in Philadelphia.
Other quantiles and the other three databases
  are shown in Table~\ref{tab:distance_quantiles}.
Figure~\ref{fig:cdf_by_city} shows the distribution of distances
  by city and database.
We use ``city" to refer to the city itself along with the 40-mile buffer around it.
Because the distance from Staten Island to North Philadelphia is only 46 miles,
  some data are included in the curves for both New York and Philadelphia.

\begin{table}
\begin{tabular}{ccccc}
{} & \multicolumn{4}{c}{New York} \\ \hline
{} & \multicolumn{2}{c}{MaxMind} & \multicolumn{2}{c}{IP2Location} \\
    Quantiles & Paid & Free & Paid & Free \\
    \hline
    0.10 &     0.73 &   0.78 &        2.96 &   3.04 \\
      0.25 &     1.37 &   1.47 &        6.06 &   6.14 \\
      0.50 &     2.62 &   2.82 &       12.04 &  12.14 \\
      0.75 &     4.95 &   5.49 &       30.05 &  30.48 \\
      0.90 &     9.66 &  11.02 &       61.84 &  63.01 \\
      \hline
{} \\
{} & \multicolumn{4}{c}{Chicago} \\ \hline
{} & \multicolumn{2}{c}{MaxMind} & \multicolumn{2}{c}{IP2Location} \\
    Quantiles & Paid & Free & Paid & Free \\
    \hline
    0.10 &    0.97 &   1.02 &        4.58 &    4.73 \\
      0.25 &    1.75 &   1.86 &        9.79 &    9.91 \\
      0.50 &    3.31 &   3.57 &       23.95 &   24.28 \\
      0.75 &    6.42 &   7.10 &       45.58 &   45.73 \\
      0.90 &   13.04 &  16.09 &      196.43 &  202.85 \\
      \hline
{} \\
{} & \multicolumn{4}{c}{Philadelphia} \\ \hline
{} & \multicolumn{2}{c}{MaxMind} & \multicolumn{2}{c}{IP2Location} \\
    Quantiles & Paid & Free & Paid & Free \\
    \hline
    0.10 &         1.13 &   1.20 &        4.16 &   4.19 \\
      0.25 &         2.14 &   2.27 &        8.97 &   9.01 \\
      0.50 &         4.02 &   4.29 &       20.87 &  20.99 \\
      0.75 &         7.57 &   8.23 &       39.60 &  39.71 \\
      0.90 &        13.53 &  15.03 &       78.34 &  78.79 \\
      \hline
\end{tabular}

\caption{Quantiles of accuracy in kilometers, for each database and city.}
\label{tab:distance_quantiles}
\end{table}

Although the paid databases are more accurate
  in each city and at every quantile,
  the relative improvements in accuracy are modest.
An important limitation of this particular study is our focus on urban areas
  in the United States.
In particular, we do not test accuracy of these databases
  outside of major metro areas, and 
  global or national performance may of course be different.
Nonetheless,
  it would be possible to perform the analysis we have presented in this
  section for other datasets, if and when they are made available.

\section{The Geography of Consumer Subnets}\label{sec:results}

We now turn from an initial assessment of the dataset and databases,
  to measurements of the geography of the underlying networks.

\subsection{Under what circumstances are IP geolocation databases accurate?}
\label{ssec:reliable}

The basic results of Section~\ref{ssec:db_error} mask extreme but unsurprising heterogeneity.
  Figure~\ref{fig:cdf_by_city} already shows that geolocation performs
  better in New York than Chicago, and better in Chicago than Philadelphia.
But the largest source of heterogeneity stems from providers,
  which deploy different physical infrastructures
  (and serve different cities).
This entire Section relies entirely on the free databases.

\begin{figure}
\centering
\includegraphics[width=0.6\linewidth]{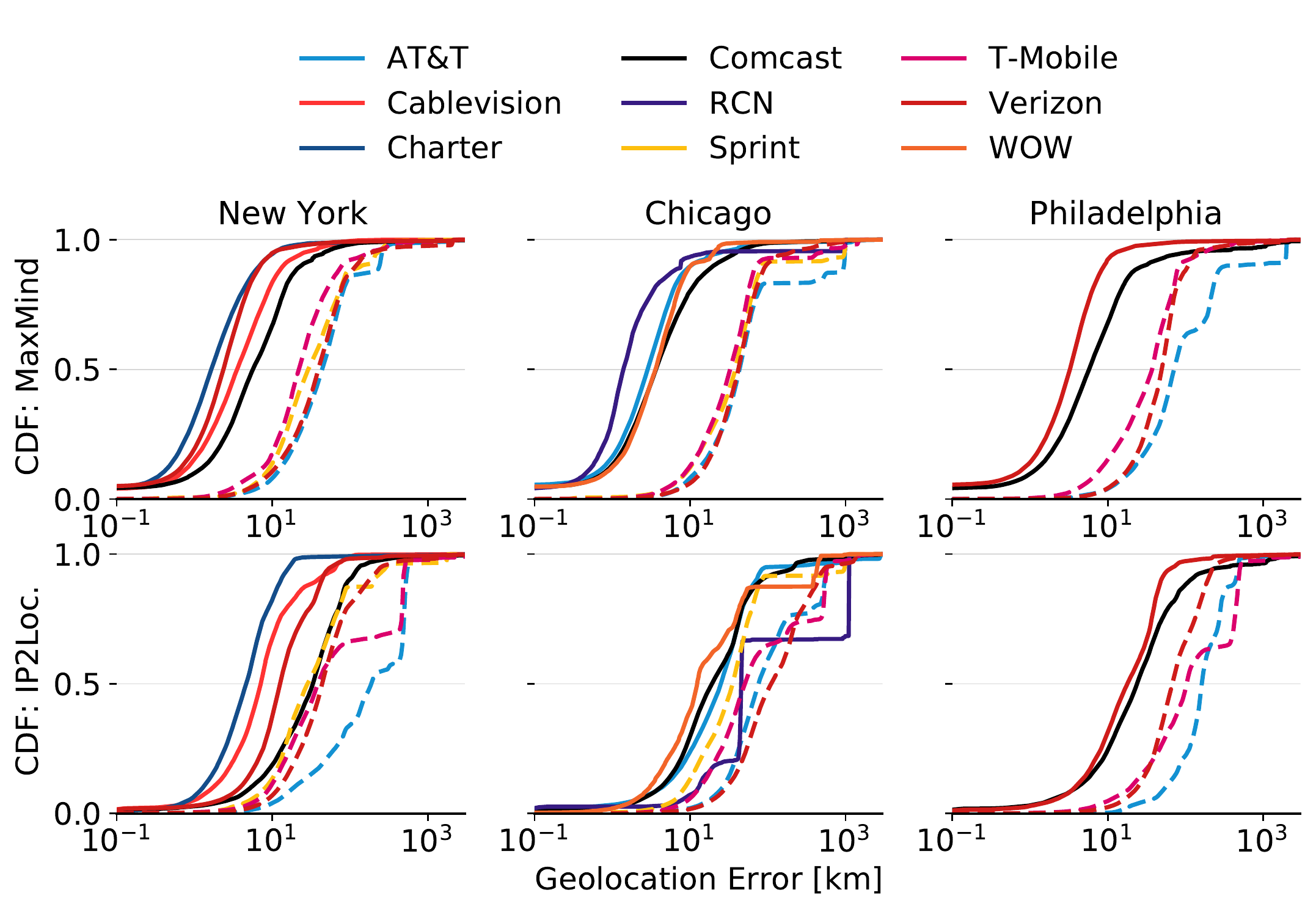}
\caption{Geolocation performance by city, database provider, and ISP. Free versions of the database are used in each case.  ISPs are shown by their ``brand" colors, according to the whois database, which leaves the Sprint and T-Mobile networks distinguishable. Fixed-line networks are denoted by solid lines while mobile networks shown by dashed lines. \label{fig:cdf_cities_dbs}}
\end{figure}

\paragraph{Fixed-line and mobile networks.}
Figure~\ref{fig:cdf_cities_dbs} shows accuracies observed in New York, Chicago, and Philadelphia
  for major broadband carriers in each market.
Again the CDF is the share of location reports.
In the best cases, such as either RCN or Comcast on MaxMind in Chicago,
the median error is less than 5~km.
In each city/database pair,
  the accuracy is good for fixed broadband
  and poor for any mobile broadband.
This Figure, and others in the main text, 
  rely on ISP classification via whois, as described in Section~\ref{ssec:unacast}.
A version of this Figure based on an IP addresses' ASNs,
  is included in the appendix, and is very consistent.

In Chicago,
  MaxMind is more accurate with fixed-line 
  (AT\&T, RCN, WOW, and Comcast) than on
  mobile (AT\&T Mobile, T-Mobile, Sprint, Verizon Mobile)
  carriers.
(IP2Location performs poorly with RCN.)
Similarly in New York, Charter, Cablevision, Comcast and Verizon are
  better localized than AT\&T Mobile, Sprint, T-Mobile, and Verizon Mobile;
  and in Philadelphia, geolocation is more accurate on Comcast than Verizon,
  T-Mobile, AT\&T Mobile, or Verizon Mobile.

Quantitatively, the share of Comcast data in New York
  that MaxMind's free service locates within 10~km of the GPS location is 67\%.
At the other extreme, 87\% of T-Mobile location reports from the New York region
  are IP geolocated to just two distinct locations representing New York itself and Newark;
  98\% are assigned either to those two, or
  to one of six other locations in Philadelphia (3), Providence, Boston, and Washington.
As a result, only 18\% of devices are assigned within 10~km of their true location.
In fairness, it must be emphasized that MaxMind does not \emph{claim} to
assign these devices within 10~km: almost all of the T-Mobile addresses assigned to the New York and Newark
locations are in the 200~km accuracy class.

This basic dichotomy between mobile and fixed broadband
is apparent even within ISPs.
AT\&T offers both services in Chicago, 
  and the CDFs for its fixed-line and mobile services are widely separated.
The individual subnets with
  the largest geolocation errors all 
  belong to the AT\&T Mobility organization.
In New York and Philadelphia,
  AT\&T only operates mobile networks,
  and this is reflected in those cumulative distributions.
The observation that mobile and fixed-line networks differ 
  may appear obvious once stated, but it need not have been true.
Mobile carriers could have constructed networks and CG-NATs
  with a fixed set of public IP addresses at each antenna.
That does not appear to be what they did.

\begin{figure}
\centering
\includegraphics[width=0.65\linewidth]{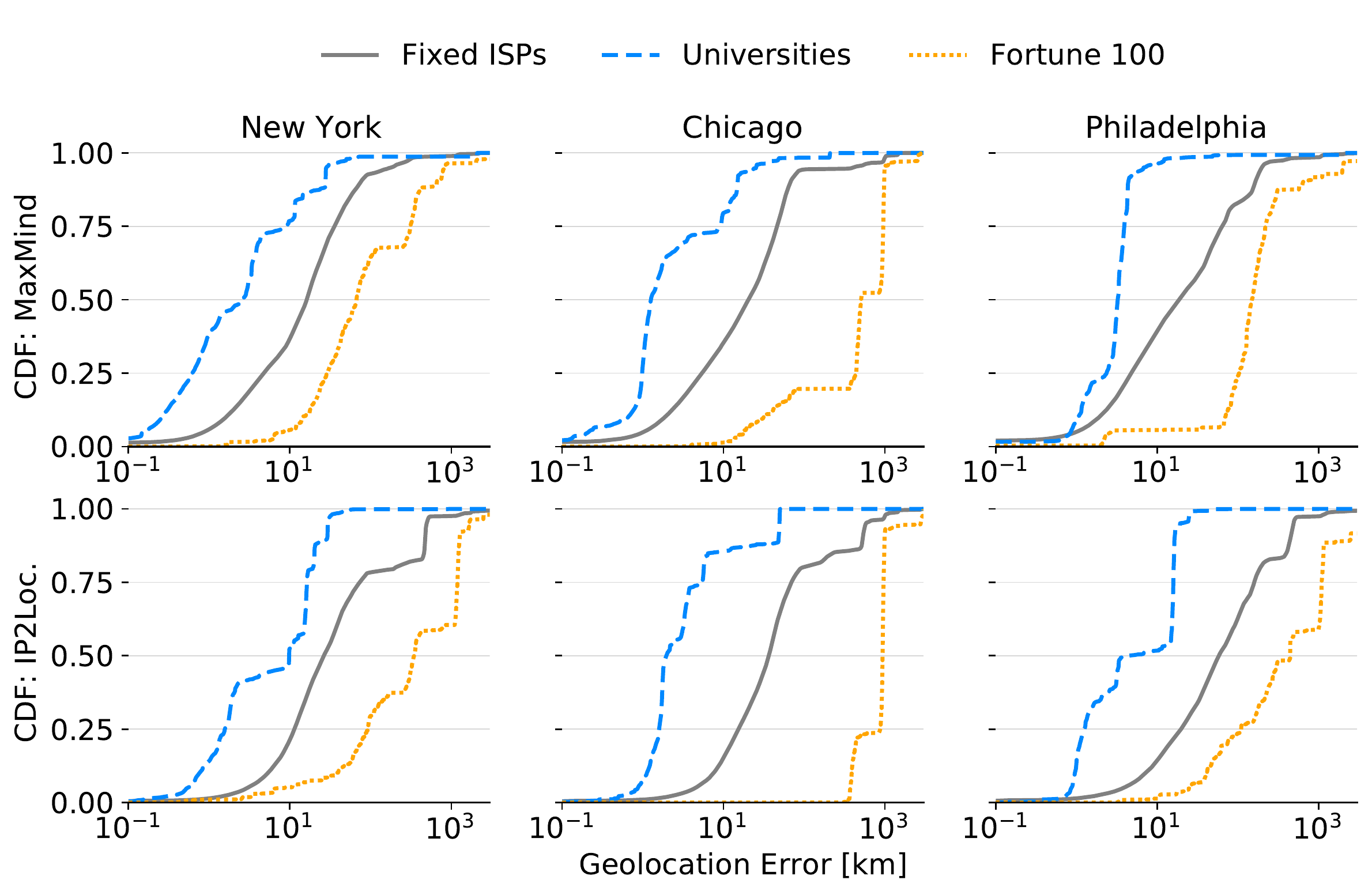}
\caption{Geolocation performance on consumer ISPs, contrasted with large universities and Fortune 100 companies. \label{fig:f100_edu}}
\end{figure}

\paragraph{Universities, businesses, and consumer networks.}
Before continuing, we also contrast geolocation performance on
consumer fixed-broadband, with large universities and companies.
We include universities with at least ten thousand students,
and Fortune 100 companies other than ISPs.
Again, we note that we are implicitly studying the Wi-Fi
  access points that these institutions operate and which
  their employees, students, and clients connect to via mobile devices,
  rather than wired connections or fixed infrastructures of servers.
Universities are a classic target in the academic literature on geolocation,
  but Figure~\ref{fig:f100_edu} shows that they are
  in general more-accurately geolocated than either
  consumer ISPs or companies.
This is not surprising: they have large, physically-concentrated networks,
  with registration addresses clearly spelled out in ARIN records.
In most cases, median geolocation error on MaxMind (free) is less than 2~km,
  though a few institutes -- DePaul in Chicago and the City University of New York --
  are mislocated by upwards of 10~km.
Note that the nominal sample period is August 2020,
  when students~-- and indeed many staff and faculty~--
  were not on campus, due to both summer vacation and the coronavirus pandemic.

Figures \ref{fig:cdf_by_city}-\ref{fig:f100_edu}
  suggest that for a substantial share of traffic,
  IP geolocation is quite accurate.
However, this does not do us much good unless those locations
can be identified in advance.
It is already clear that the picture is rosier with fixed broadband.
Those data can be easily identified,
either via a \texttt{whois} look-up or (in some cases)
through the geolocation databases themselves.
But mobile and fixed is not the only lever.
MaxMind is able to perform better on RCN than on Comcast in Chicago,
and better on Charter or Cablevision than Comcast in New York.
How are we to identify localizable blocks of addresses?

\begin{figure}
\centering
\includegraphics[width=0.45\linewidth]{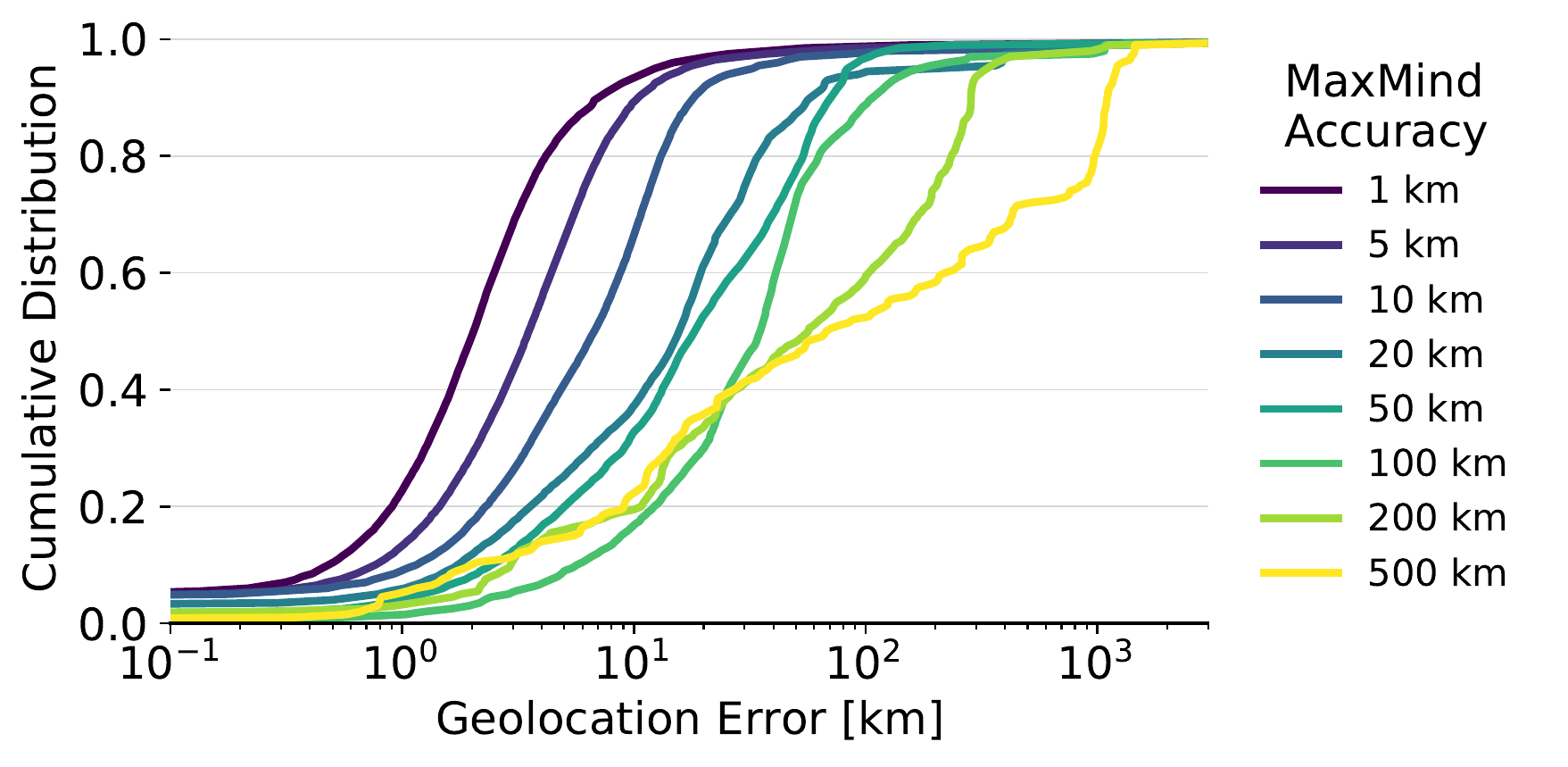}
\caption{Cumulative distribution of geolocation accuracy on the MaxMind database, by quoted accuracy bin. \label{fig:cdf_mm_by_accuracy}}
\end{figure}

We highlight two additional methods.
MaxMind's database provides an ``accuracy'' field that successfully
identifies the precision of entries.
Figure~\ref{fig:cdf_mm_by_accuracy} shows the CDF for
successive bins of claimed accuracy on the free database.
In the most precise bin, accuracy of ``1~km,''
the median device in Chicago is geolocated just 2.0~km
from the GPS-based location.
The ``error'' with respect to the ground-truth
degrades in-line with quoted accuracy,
though there is enormous spread in the least-precise, 500~km bin.
It is thus \emph{possible} to identify
accurately-located addresses -- MaxMind does it.
But this leaves an open question:
\emph{why} are those addresses well or ill-located?

That brings us to the second method.
Our hypothesis is that
if /24 subnets are geographically localized~-- small~--
then addresses within them are more-likely to be accurately geolocated.
If they are large, then precise locations would require finer address-level data.
The question can then be re-posed:
how large are subnets, and is their size in fact correlated with geolocation accuracy?

\subsection{What is the geographic scale of /24 subnets?}
\label{ssec:size}

What are the physical and network properties of accurately-located subnets?
We focus this analysis on a single, fixed network~-- Comcast~--
  and require that subnets have at least 10 devices and 10 distinct IP addresses.
There are over twenty thousand such subnets between the three cities.

\begin{figure}
\centering
\includegraphics[width=0.42\linewidth]{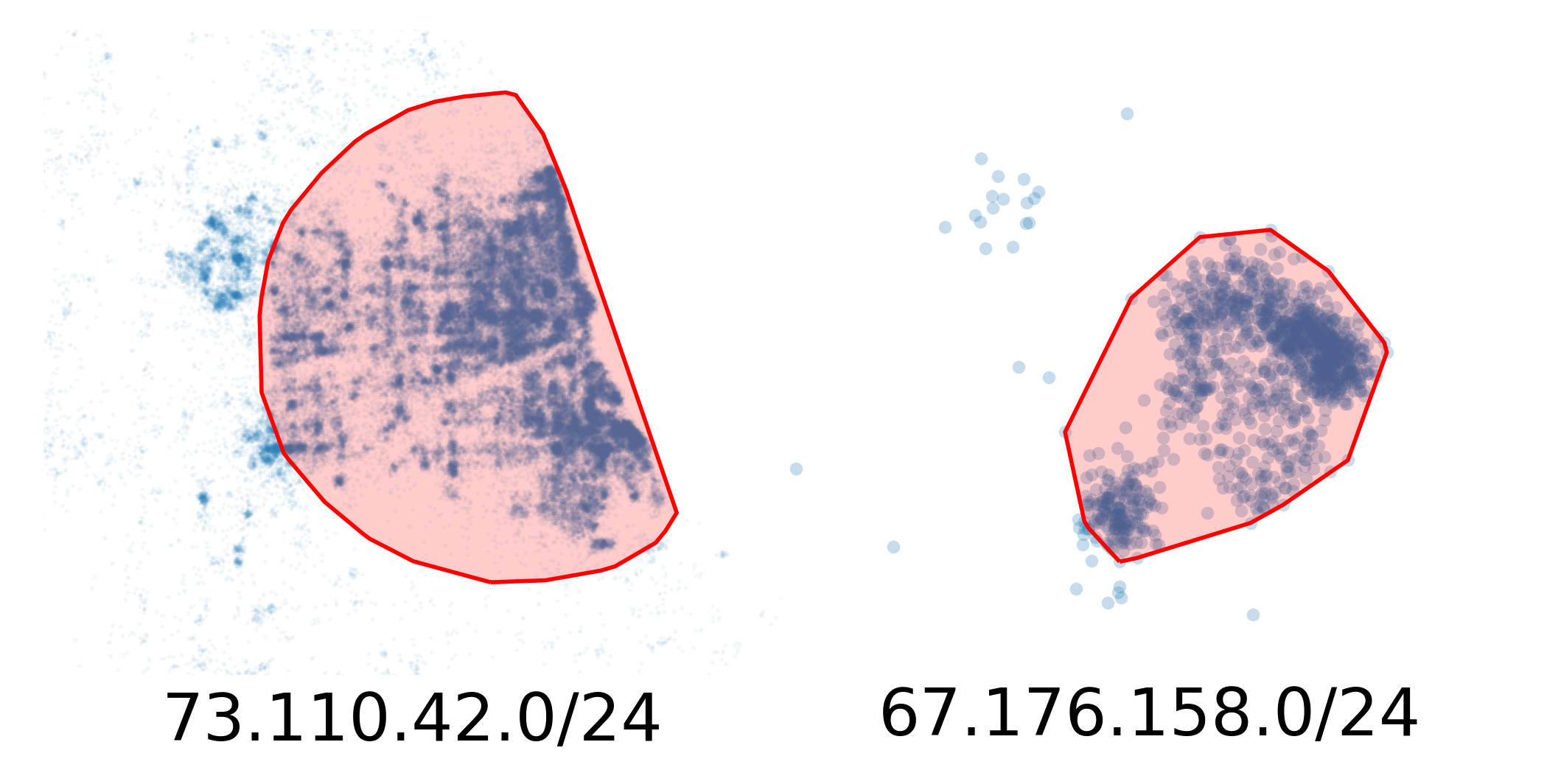}
\caption{Illustration of the procedure defining subnet scales, for one dispersed and one well-localized subnet in Chicago.  Convex hulls wrap around $f = 0.9$ of the points within the subnet.  The ``scale" is the square root of this area. The linear scale on the right-hand side (67.176.158.0/24) is a factor of 8 larger than on the right-hand side. Gaussian noise has been added to the locations for illustrative purposes only. \label{fig:convex_hull}}
\end{figure}

\paragraph{Constructing a physical scale.}
To quantify whether or not a subnet is localized,
we define a characteristic physical scale.
Many subnets have some outliers,
perhaps with locations reported after the fact.
To mitigate the impact of these outliers,
we must first identify them.
We compute the medioid of locations in the subnet,
defined in this case simply as the median of the $x$ and $y$ coordinates
in a projected (flat) geometry (EPSG 2163).
We then measure individual locations' distances from that medioid.
We select a configurable fraction $f$ of the data that is ``closest'' by that measure.
For that subset of the data, we calculate the convex hull.
If $f = 1$, then the convex hull covers all locations recorded on the subnet;
if $f = 1/2$, it covers the half of points closest to the medioid.
Finally we take the area of the convex hull,
and ``convert'' this area to a distance by taking its square root.
That square root defines the length scale of the subnet.
Figure~\ref{fig:convex_hull} illustrates this procedure for two subnets.
(To preserve anonymity,
random noise has been added to the individual points in the illustration.)

\begin{figure}
\centering
\includegraphics[width=0.4\linewidth]{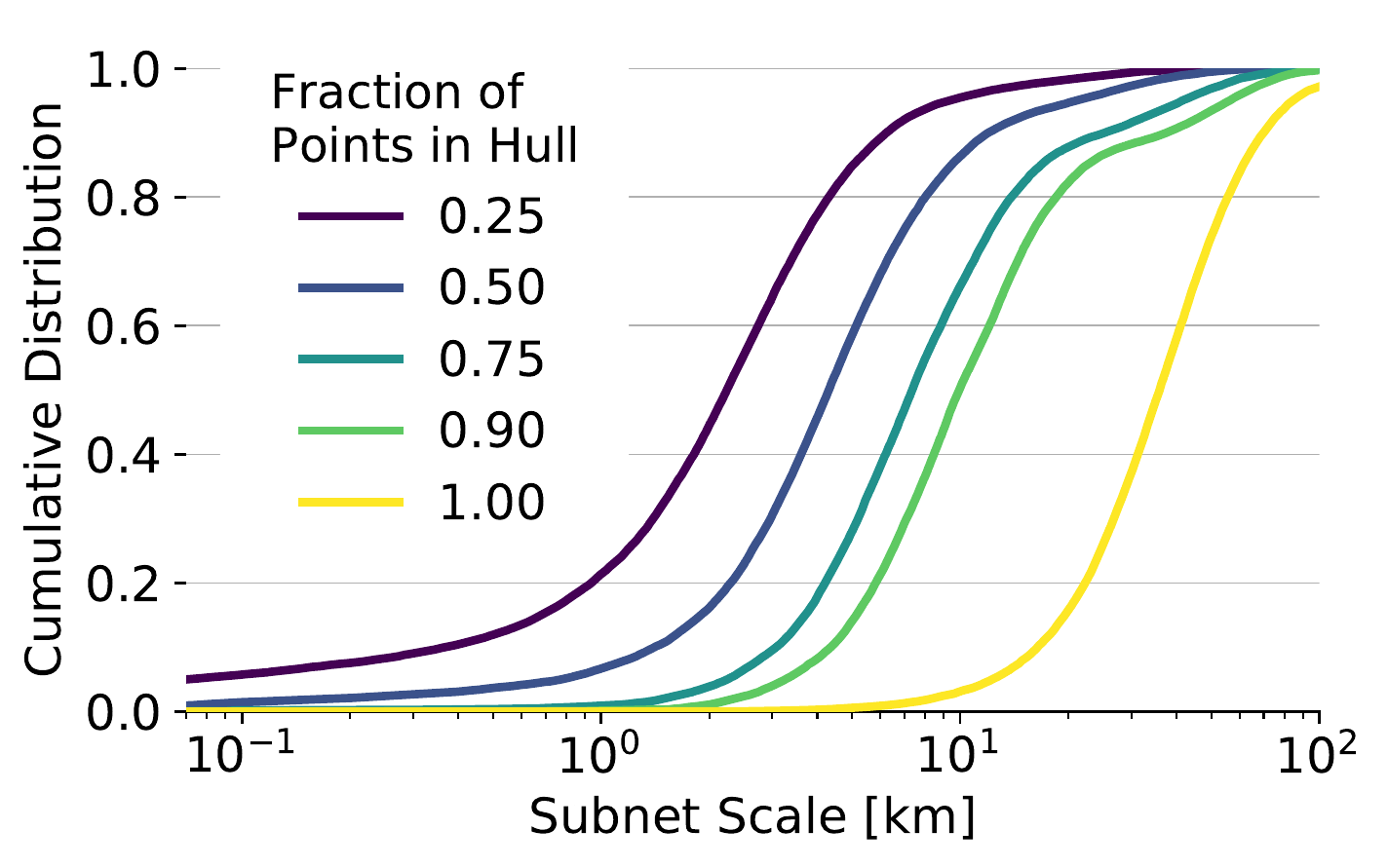}
\caption{Cumulative density of subnets' distance scale as derived from the convex hull of locations, as described in the text. \label{fig:subnet_scale}}
\end{figure}

Figure~\ref{fig:subnet_scale} shows this distance scale
  for subnets with at least 10 devices and addresses,
  for several choices of $f$.
By construction, 
  the scale is smaller or larger when outliers are more or less suppressed,
  respectively.
Setting $f = 0.5$ results in a median subnet scale of 4.3~km,
  and $f = 0.9$ leads to a scale of 9.9~km.
However,
  the proportion of subnets with scales exceeding 10~km is small
  for any choice of $f < 0.9$.

\paragraph{The relationship of physical scale and accuracy.}
Armed with this scale, we return to the earlier question:
when can subnets be accurately located?
Discarding locations with geolocation error over 100~km,
the correlation between $f = 0.75$ subnet scale and mean geolocation error,
  is 0.69 for MaxMind Free (GeoLite) 
  but only 0.30 on IP2Location (which has worse overall performance).
We thus confirm
  the hypothesis that localization and localizability are related,
  though strictly speaking, this analysis is not causal.

Still, this analysis has delayed
  but not \emph{answered} the question;
  it suggests that geolocation fails on fixed-line addresses when subnets are large,
  which raises in turn the issue of why large subnets exist at all.
Comcast uses both large and small subnets.
Are large ones used differently?

\begin{figure}
\centering
\includegraphics[width=0.4\linewidth]{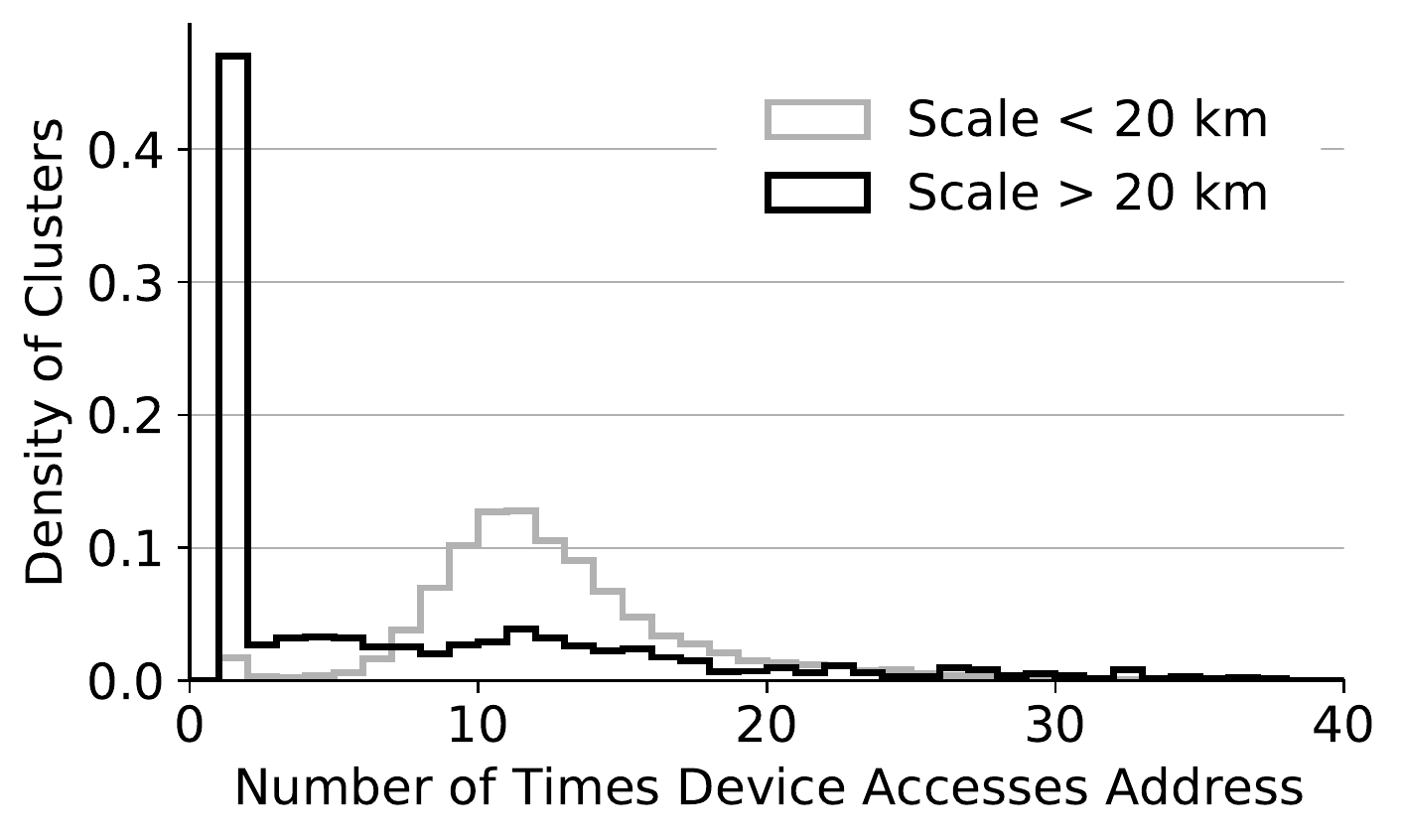}
\caption{The number of times a single device visits a single IP address on the subnet (weighted by visits).  On subnets with scale greater than 20~km ($f = 0.75$), nearly half of visits device/IP pairs are unique. \label{fig:address_visits}}
\end{figure}

We hypothesize that the spatially-concentrated subnets are nearly static
  whereas large ones provide a reserve of ``ephemeral" addresses ~--
  perhaps for devices awaiting assignment of a long-term address.
A client assigned to an ``ephemeral" address
  would be unlikely to fall on that same address again,
  whereas a ``sticky" address granted to a home network would be used repeatedly.
The relevant variable is thus the number of times
  that a single client is observed at each IP address
  (weighted by visits).
Figure~\ref{fig:address_visits}
  confirms the hypothesis:
  for subnets with scale greater than 20~km ($f = 0.75$)
  nearly half of visitors to an IP address visit exactly once.

\subsection{How persistent are the physical locations of /24 subnets?}
\label{ssec:movement}

Geolocation providers are quick to point out that
  databases evolve continuously.
Clearly, the physical infrastructure of the Internet evolves over time,
  but how quickly do subnets actually move?
Because mobile networks subnets are already physically
  very large, and addresses on them are not accurately located,
  we focus this analysis on fixed-line broadband.

\begin{figure}
\centering
\includegraphics[width=0.4\linewidth]{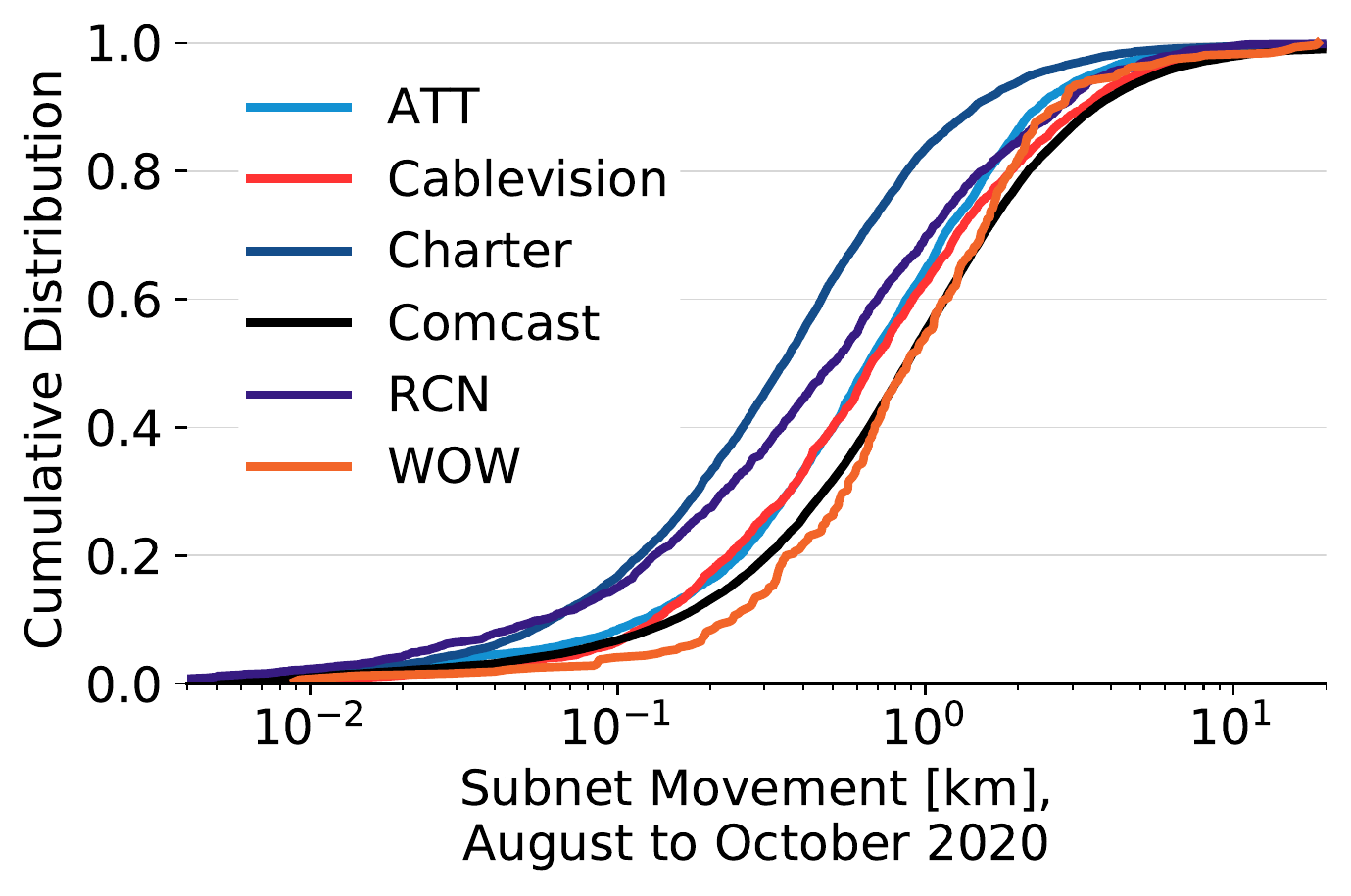}
\caption{Distance moved by the medioids of /24 subnet on fixed-line networks, over a two-month period from August to October 2020. \label{fig:movement_by_month}}
\end{figure}

\paragraph{The movement of subnets.}
Figure~\ref{fig:movement_by_month} presents
  the physical distance between the medioids of
  individual /24 subnets, as constructed in August and October 2020.
As in Section~\ref{ssec:size}, the medioid is the median of the $x$ and $y$ coordinates.
To enter into this figure, subnets must have at least ten unique devices
  and ten unique addresses in each month.
We consider only fixed-line broadband carriers, for this exercise.

On each network considered, 
  the median subnet moves less than a kilometer;
There is some inherent variability in our construction
  of the medioid as the ``location'' of the subnet in each period,
  and the Figure shows the difference of these two ``noisy" measurements.
We thus suspect that this overstates movement.
In short, we conclude that on this time scale,
  subnet locations are quite stable

\paragraph{Is the sample biased?}
A substantial threat to this analysis is sample composition:
by requiring 10 devices and 10 addresses,
the subnet \emph{must} be observed in New York, Chicago, or Philadelphia
in both months, to enter the sample at all.
However, it does not seem to be the case that subnets are moving out of sample.
Of the subnets satisfying the cuts in August, 92\% also pass them in October (vice versa, 96\%).
If we raise the thresholds to enter the sample, requiring 20 devices and 20 addresses,
95\% of subnets passing these cuts in August also show up with at least 10 devices in October (vice versa, 98\%).
Raising the thresholds yet further to 50 devices and 50 addresses,
the persistence from August to October exceeds 99\% (vice versa, 98\%).

\subsection{How long does a consumer connection retain an IP address?}
\label{ssec:churn}

The analyses above show that 
  IP addresses identify physical locations
  at the level of 2~km, under the best circumstances.
On its own, the IP address clearly does not identify individuals.

Of course, physical locations~-- geographic coordinates~-- 
  are not the only way in which IP addresses
  identify people.
Linked to log-ins or other online behaviors,
  IP addresses can be used to track users over time
  even without cookies or fingerprinting (or as a component of a fingerprint).
If the IP address is static for a long time, 
  it easier to link online behaviors.
A critical concern is thus \emph{how long}
  fixed-line IP addresses remain with a single household.

\paragraph{Defining churn.}
We define \emph{churn} as the likelihood 
  of a device returning to the same IP address on an ISP,
  after a delay of $d$ days.
The denominator includes every pair 
  of night-time connections by a single device to one ISP,
  $d$ days apart.
We select night-time activity, 
  to focus on periods when devices can be reasonably assumed ``at home."
The numerator is the number of those
  pairs for which the two nights' connections
  are on the same IP address.
Stated less formally:
  if I see a device on Monday night ($d = 0$)
  and again on the same ISP Tuesday night ($d = 1$),
  what are the chances that it will be on the same IP address?
What about next Monday ($d = 7$)?

Since the sample selection is somewhat peculiar~--
  devices are necessarily recorded on fixed-line broadband on multiple nights~--
  one should take some care in interpreting these results.
This consideration is particularly acute at the maximum of the range,
  since there are fewer opportunities
  for a device to be observed 80 days apart (just 10)
  let alone 90 (just 1).
This perhaps explains the drop-off on the right-hand side.

\begin{figure}
\centering
\includegraphics[width=0.6\linewidth]{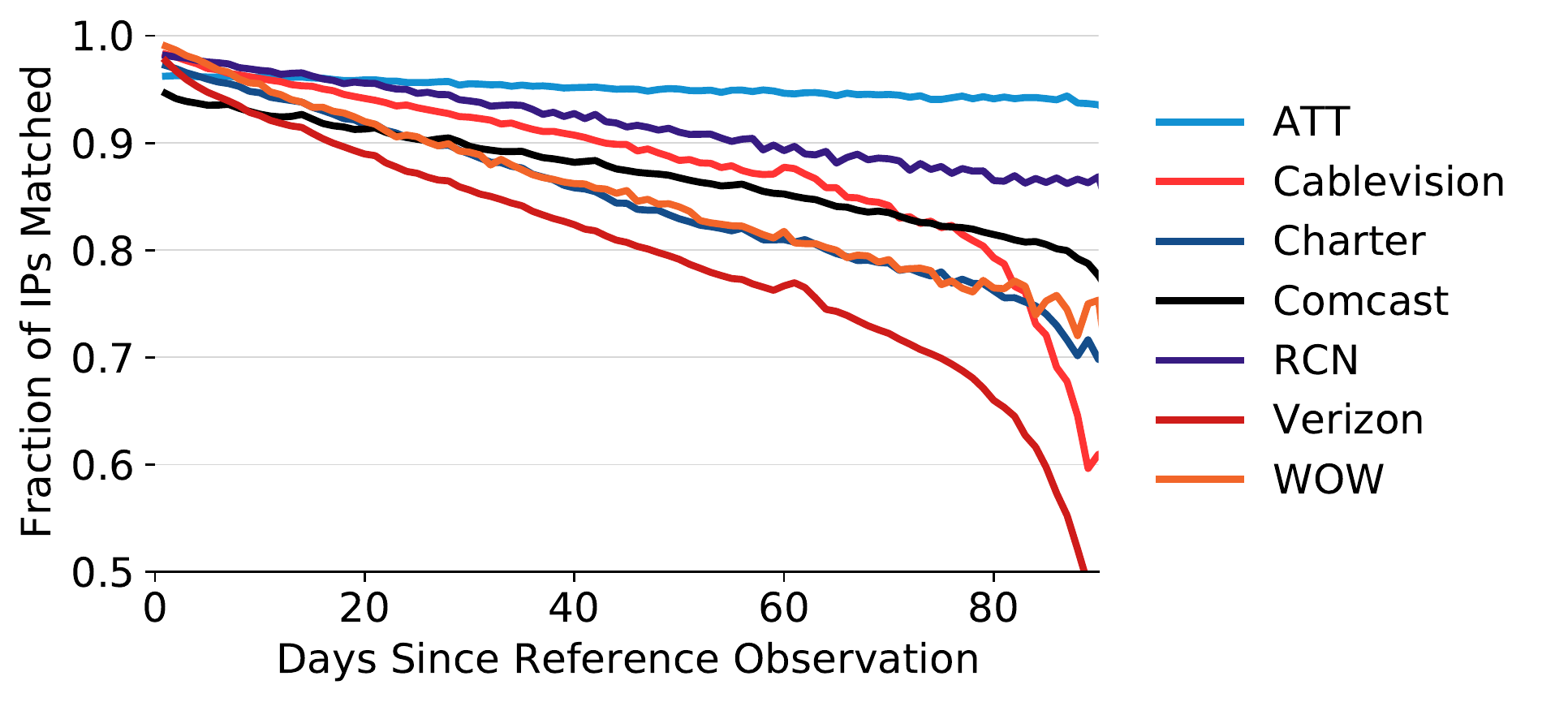}
\caption{Persistence of IP addresses.  The Figure shows the share of night-time clusters on a single ISP and device, separated by $d$ days, for which the IP addresses are equal on both clusters.  Note that for visual clarity, the $y$ axis begins at 0.5 instead of 0.
\label{fig:ip_stability}}
\end{figure}

\paragraph{Rates of change, over two months.}
Figure~\ref{fig:ip_stability} shows the persistence of 
  IP-addresses on fixed-line broadband ISPs.
It is clear that devices ``leave" individual 
  IP addresses gradually, but at different rates on different ISPs.
After one month, more than 90\% of
  devices observed reconnecting to AT\&T, RCN, and Cablevision
  do so on the same IP address.
After two months,
  more than three-quarters of devices return to the same IP address,
  for all major ISPs in the three cities shown.

\section{Can IP geolocation databases be used to study Internet access?}
\label{sec:discussion}

At this stage, we would usually turn to a general discussion of findings.
Here, we focus our discussion and extend our results,
  according to the question that originally motivated our work:
  assessing the potential for using IP-referenced data
  in \emph{social science} research on Internet access.
\emph{Where} and \emph{for what demographic groups} is geolocation accurate?
Can \emph{IP geography} enable \emph{Internet demography}?
To make this query concrete, imagine a study of the ``homework gap" --
  (in)equity in access to digital resources for education --
  based solely on server logs from a site like Wikipedia.
If we observe frequencies of use by IP subnet \emph{alone},
can we infer what groups do and do not access the site?

\paragraph{General considerations.}
This question is non-trivial,
since it confronts the correlations of
population density and demographics with geolocation accuracy,
along with the spatial patterns of connection modality (mobile vs fixed).
Cities have smaller subnets simply
because they have higher density of people and devices.
They also tend to have larger minority populations.
This alone leads to a correlation between geolocation accuracy with
demographics or disadvantage.
For Chicago and its buffer,
  the correlation between tract
  median geolocation error on MaxMind (free) 
  and population density is $-0.09$ ($p < 0.0001$);
in turn, population density is correlated with log median household income
($r = -0.18$, $p < 10^{-10}$).
Both of these are small but significant.
The flip side of better accuracy at higher density
is that distance precision \emph{has} to improve
in dense environments, to associate activity with the right population.
It's easier to ``jump" over many people when they are close together.

Accuracy also varies \emph{within} the city,
due to heterogeneity in the fraction of
people on mobile vs fixed broadband.
There are two reasons for this.
People use mobile devices (1) when they are on the go,
or (2) because they do not have access to a fixed broadband connection at home.
That means that devices in the present sample observed in city centers
appear to have ``inaccurate" IP geolocation,
simply because the device users are more-likely on mobile on the way to or at work.  % \fnote{Appendix: Add table of MaxMind accuracy at all times (to show daily trends)?}
On the other hand,
populations without fixed broadband access
are unlikely to be accurately IP geolocated, even in their home neighborhood.

As a final consideration before proceeding,
one must not confound ``unknown" addresses with ``mis-located" ones.
For example, if a default database location for T-Mobile addresses sits in a particular neighborhood,
that neighborhood will appear to have ``accurate" geolocation, even though
the locations are not known any better than elsewhere.
Performance will appear to ``degrade" radially, with distance from the default location.
Since the default locations are usually in or near cities,
that would (\emph{ceteris paribus}) give a false impression
  that IP addresses in cities (or near the center of the United States, for instance)
  are accurately-located. % \cite{2016_arstechnica_kansas,2019_gizmodo_south_africa}

\begin{figure}
\centering
\includegraphics[width=0.4\linewidth]{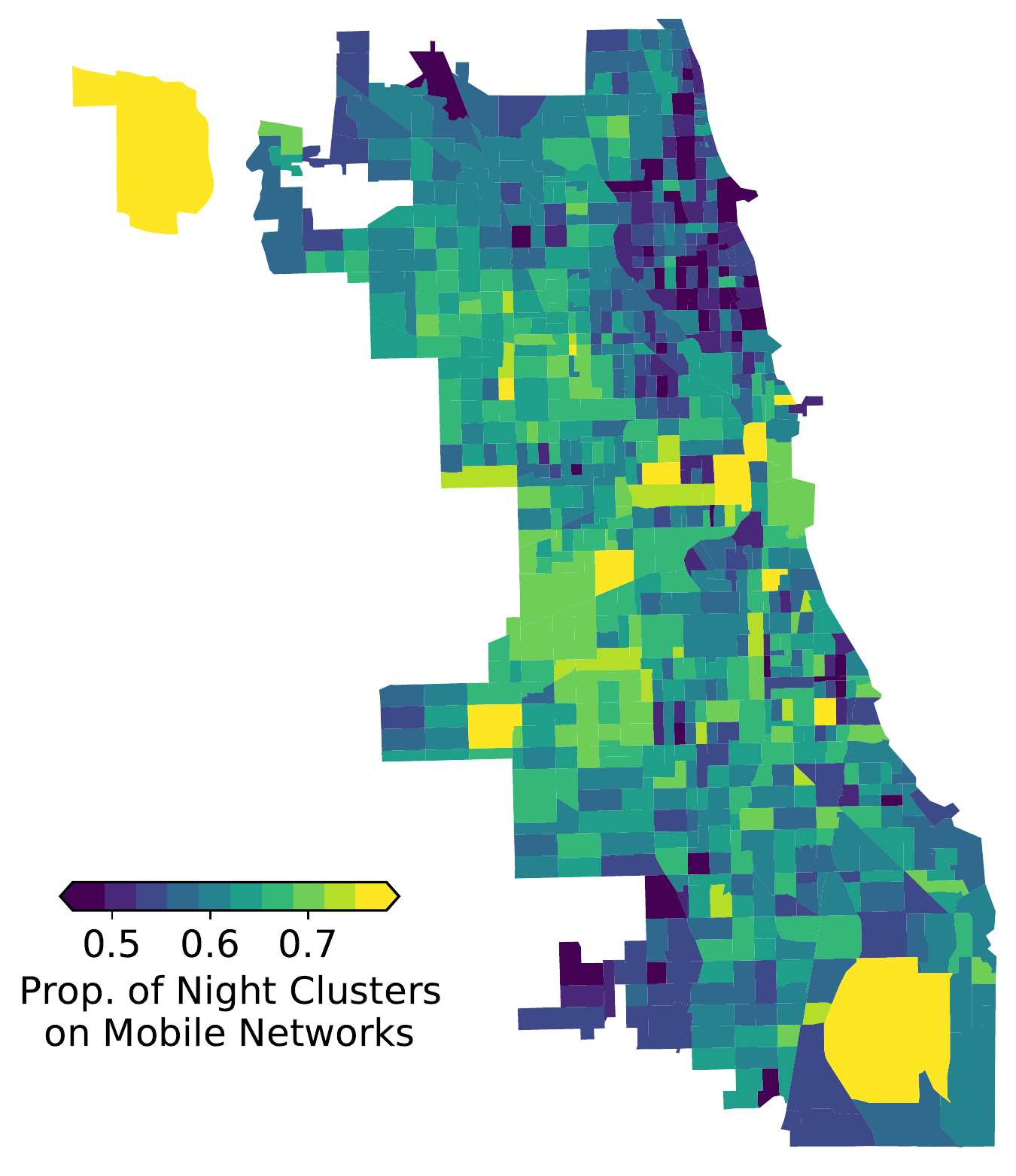}
\caption{Proportion of night-time clusters in Chicago recorded on mobile networks. \label{fig:chicago_mobile}}
\end{figure}

\paragraph{Differences in access modality by demographic group.}
Returning to the data, Figure~\ref{fig:chicago_mobile} presents the proportion of the night-time clusters in
each tract of Chicago, that are on fixed and mobile broadband.
Note that the data are inherently mobile devices with GPS chips;
this does not include laptops, for instance.
This classifies AT\&T, Comcast, WOW, and RCN, as fixed-line providers,
  and T-Mobile, Sprint, Verizon, and AT\&T Mobile as mobile.
For those familiar with Chicago, the results are no surprise:
the proportion of night-time pings on mobile networks
is lower on the wealthier North Side of the city than on the West or South Sides.
Indeed, our eyes do not deceive us:
the tract level correlation between this constructed variable
and share of households with a broadband contract as reported to the Census is $-0.25$.
The correlation to neighborhood proportion Hispanic is $0.23$ (both $p < 10^{-10}$).
In other words, connection type is correlated with demographic factors and broadband adoption.
This would be reflected in geolocation accuracy.
In practice,
  this means that constraining an analysis
  to accurately-located, fixed-line IP addresses
  would tend to exclude vulnerable populations from the analysis.

\begin{figure}
\centering
\includegraphics[width=\linewidth]{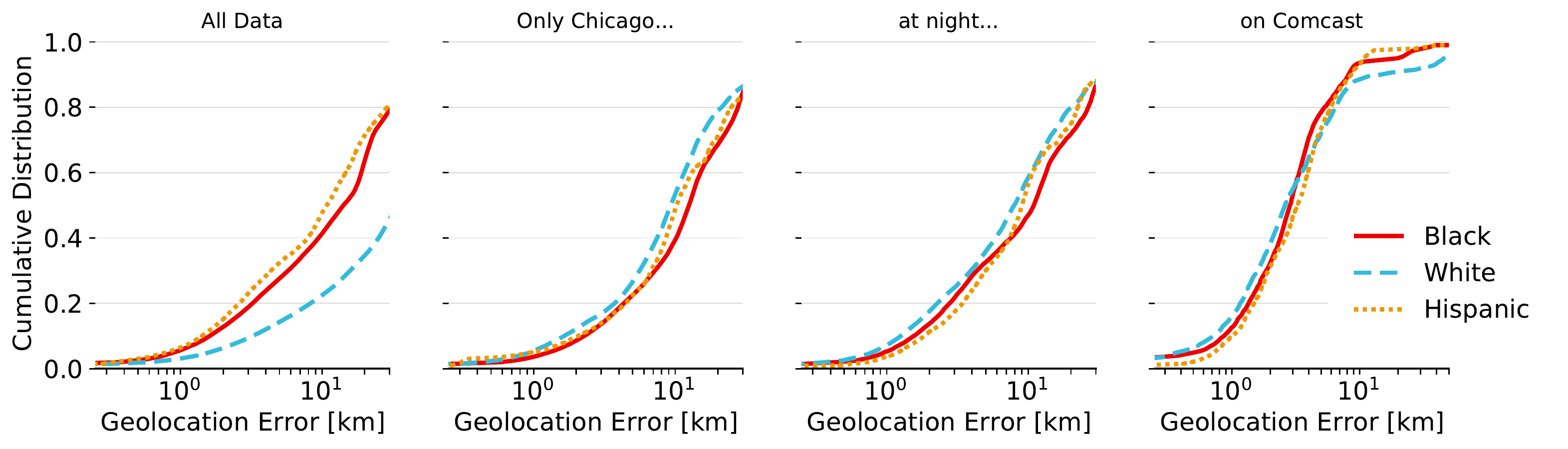}
\caption{Cumulative distribution of geolocation error for tracts with white, Black, and Hispanic super-majorities. The first panel presents all data, while the second through fourth restrict to Chicago, Chicago at night, and Chicago at night on Comcast. \label{fig:race_ethn_cdf}}
\end{figure}

\paragraph{The influences of density, demographics, and modality on IP geolocation accuracy.}
Figure~\ref{fig:race_ethn_cdf}
offers an alternative view of this effect,
disentangling the countervailing forces of density,
demographics, and access modality.
It displays the CDF geolocation accuracy
  in hyper-segregated neighborhoods of Chicago~--
ones where two-thirds of residents are white (only), Black (alone or in combination with other races), or Hispanic (of any race).
Moving from left to right,
we begin from the full dataset and layer the cumulative
requirements of devices in Chicago proper (not the 40 mile buffer),
at night (that is, likely at home), and
on Comcast (i.e., on a single, fixed broadband network).
The CDF shows the share of location reports.
The first plot shows an enormous difference between geolocation in ``white" tracts and other segregated tracts --
geolocation performs much worse.
This effect appears to have more to do with density than race:
it reverses when focusing on the City of Chicago,
and zeroing in on a single network, the performance lines up quite closely.
The exception is at the very high end (above 10~km and 90\% of the CDF),
where there is apparently an error for locations reported from ``white'' tracts.
About 80\% of points are within 5~km of the true location, for all three categories of neighborhood.

\paragraph{Attenuation bias, from reliance on mis-attributed IP addresses.}
The analyses of device modalities above
  suggest that IP geolocation databases'
  ability to attribute online behaviors to populations
  will tend to fail more often for disadvantaged groups.
Still, if we were to persist, what errors might we expect to ``accrue," by moving an
observation from its GPS-based location to the IP-based location?
In essence, this question pits the scale of geolocation accuracy against
the physical scale of demographic segregation.
If IP geolocation moves a point among communities with similar
demographics, the error does not directly bias results.

\begin{figure}
\centering
\includegraphics[width=0.5\linewidth]{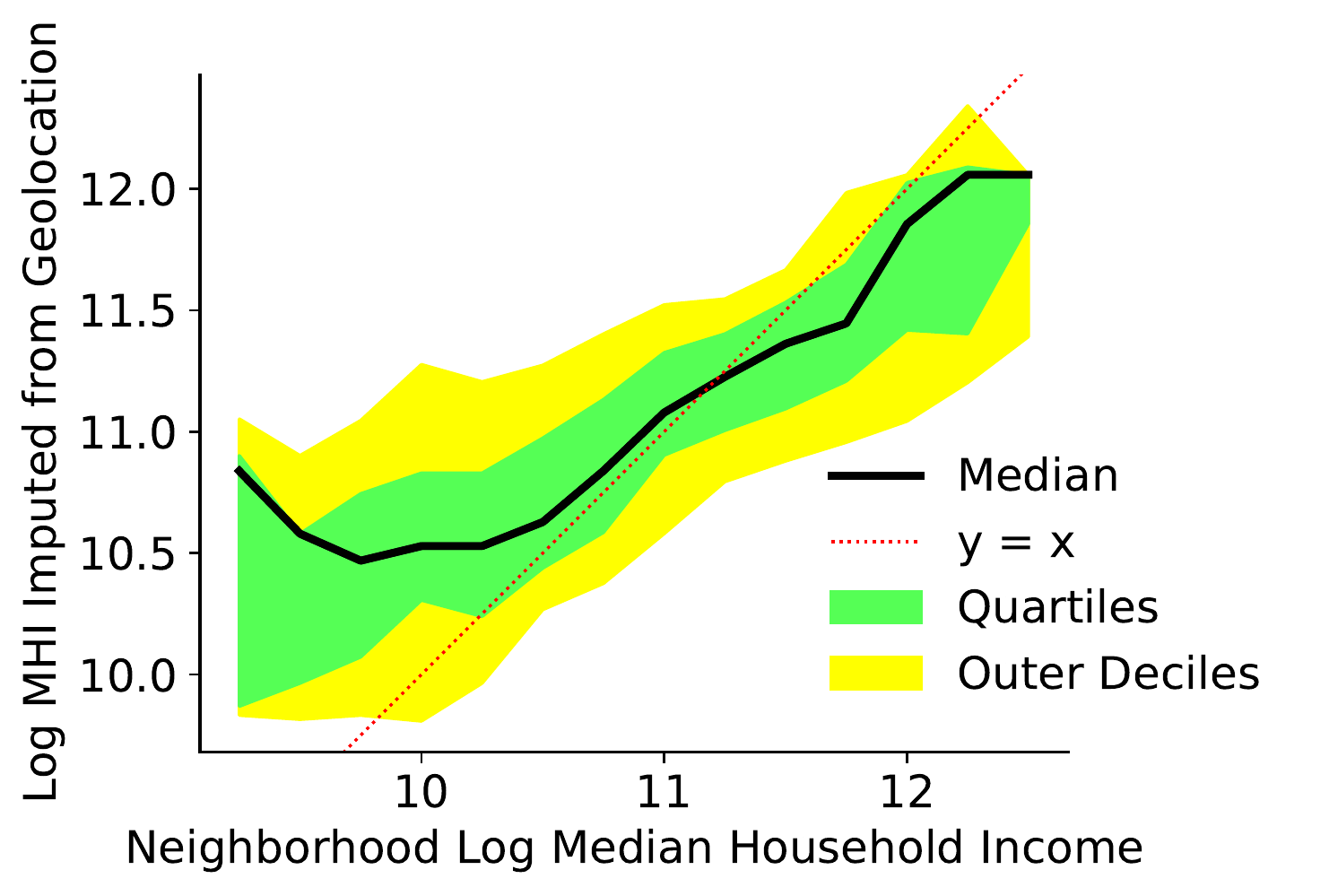}
\caption{Quantiles of neighborhood log median household income as ``imputed'' from MaxMind geolocation ($y$) as a function of the true neighborhood value ($x$). \label{fig:log_mhi_quantiles}}
\end{figure}

This illustrative analysis is limited to fixed-broadband data from Comcast,
where geolocation has a chance of succeeding.
Figure~\ref{fig:log_mhi_quantiles} presents the log median household income as it would be imputed from a MaxMind look-up,
against the true median household income of the neighborhood (Census tract).
This results in an unsurprising regression to the mean:
as is the usual case with measurement error, the slope is simply attenuated.
This suggests that even for fixed broadband,
  efforts to use IP address alone to ``link" online behaviors 
  with human populations are inadvisable at this physical scale.
They will in general yield estimates whose magnitudes are biased down.
In other words, measurements of ``who uses what"
  that rely on IP geolocation will tend to \emph{understate} differential access.
This is consistent with \citeauthor{2019_ganelin_chuang_ip_mooc_regressive}'s 
  work on the socioeconomic status of MOOC registrants.
They found that using IP geolocation to identify users' neighborhoods
  led to underestimates of inequity in adoption~\cite{2019_ganelin_chuang_ip_mooc_regressive}.

% median_error_by_geoid:1.0 Median geolocation error on MaxMind, by Census Tract. \label{fig:median_error_by_geoid}

\balance
\section{Conclusion}\label{sec:conclusion}

Using a large sample of GPS-based smartphone locations
  this paper has quantified the performance of commercial geolocation databases
  in New York, Chicago, and Philadelphia.
The precision of this analysis far outstrips past work.
The analysis has demonstrated significant heterogeneity in geolocation accuracy.
The median error for MaxMind's free service is well less than 10~km
  on fixed commercial broadband networks and at Universities.
On mobile networks, IP geolocation is not accurate below the city level.
While we consider that consumer devices in large cities in the United States
  represents a particularly useful vantage point, 
  our conclusions concerning database accuracy and and network structure
  are necessarily limited to the setting that we have observed.

Our analysis has also sought to explain \emph{why} some addresses
are accurately located whereas others are not.
The physical size of subnets is strongly correlated with accuracy.
Large subnets appear to be used for ``ephemeral" address assignment,
which clients do not use repeatedly.

Finally, we have contextualized these findings for applications to research on human populations.
Both the present data and existing surveys show that
disadvantaged populations are less likely to use a fixed broadband subscription at home.
Online behaviors cannot be accurately associated with these groups,
and dropping mobile devices altogether will tend to remove them from analyses.
Focussing on the context where geolocation does work -- fixed-line broadband~--
the accuracy still appears inadequate for associating online activities
with real-world geographies and demographics.
From a privacy perspective, a single IP address does not identify an individual,
  but it both localizes and individual and provides
  an ``index" through time that may be used to aggregate other indirect identifiers.
We have shown that the time for IP reassignment
  of fixed-line broadband consumers varies by ISP,
  but is typically on the order of months.

\newpage
\bibliographystyle{ACM-Reference-Format}
\balance\bibliography{sources}

\clearpage
\appendix
\section{Additional Plots and Tables}

\FloatBarrier

\setcounter{figure}{0}
\renewcommand\thefigure{A.\arabic{figure}}

\setcounter{table}{0}
\renewcommand\thetable{A.\arabic{table}}

\begin{table}[h!]
\centering
\begin{tabular}{lrr}
\hline
        Cluster Class &  Bumps &  Clusters \\
\hline
      Long Area Dwell &  0.382 &     0.079 \\
               Travel &  0.322 &     0.264 \\
           Area Dwell &  0.193 &     0.173 \\
     Short Area Dwell &  0.074 &     0.290 \\
 Potential Area Dwell &  0.025 &     0.127 \\
                 Ping &  0.003 &     0.066 \\
       Large Variance &  0.001 &     0.000 \\
               Moving &  0.000 &     0.000 \\
                Split &  0.000 &     0.000 \\
\hline
\end{tabular}
\caption{Proportion of bumps and clusters according to the classification type assigned by Unacast (cf Section 3.1). 
  \label{tab:unacast_classification_freq}}
\end{table}

\begin{table}[h!]
\centering
\begin{tabular}{cr}
\hline
     NIC &    Frac. \\
\hline
 AFRINIC &  0.00011 \\
   APNIC &  0.00015 \\
  LACNIC &  0.00016 \\
    RIPE &  0.00169 \\
\hline
\end{tabular}
\caption{Proportion of clusters with IP addresses in foreign Internet registries (cf Section 3.1). \label{tab:foreign_nic_fracs}}
\end{table}

\begin{figure}[h!]
\centering
\includegraphics[width=0.4\linewidth]{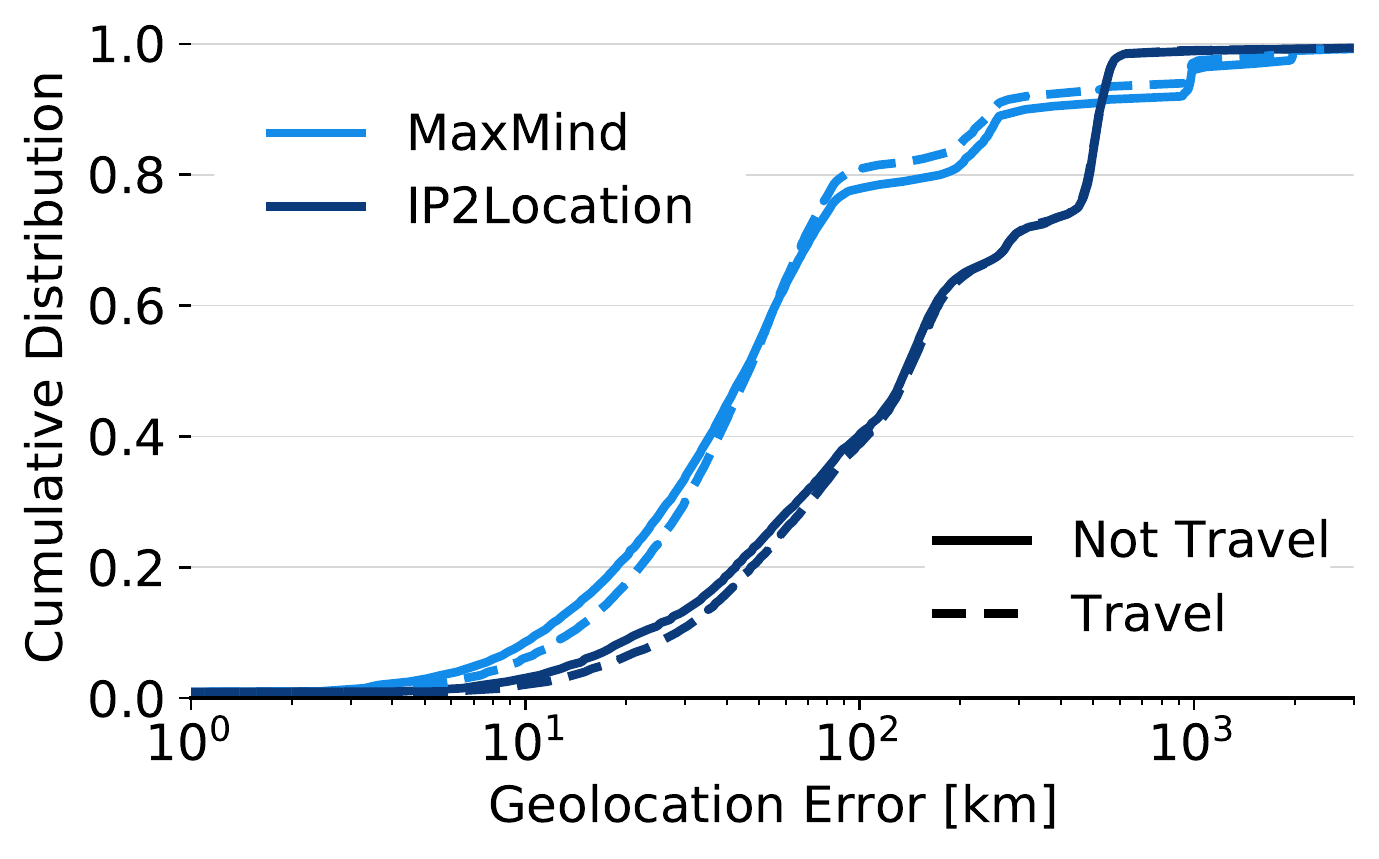}
\caption{Empirical cumulative distribution of geolocation accuracy, 
           for travel and non-travel clusters on AT\&T's mobile network, 
           as evaluated on the free versions 
             of the MaxMind and IP2Location databases (cf Section~4.1).\label{fig:travel_no_travel}}
\end{figure}

\begin{figure}
\centering
\includegraphics[width=0.6\linewidth]{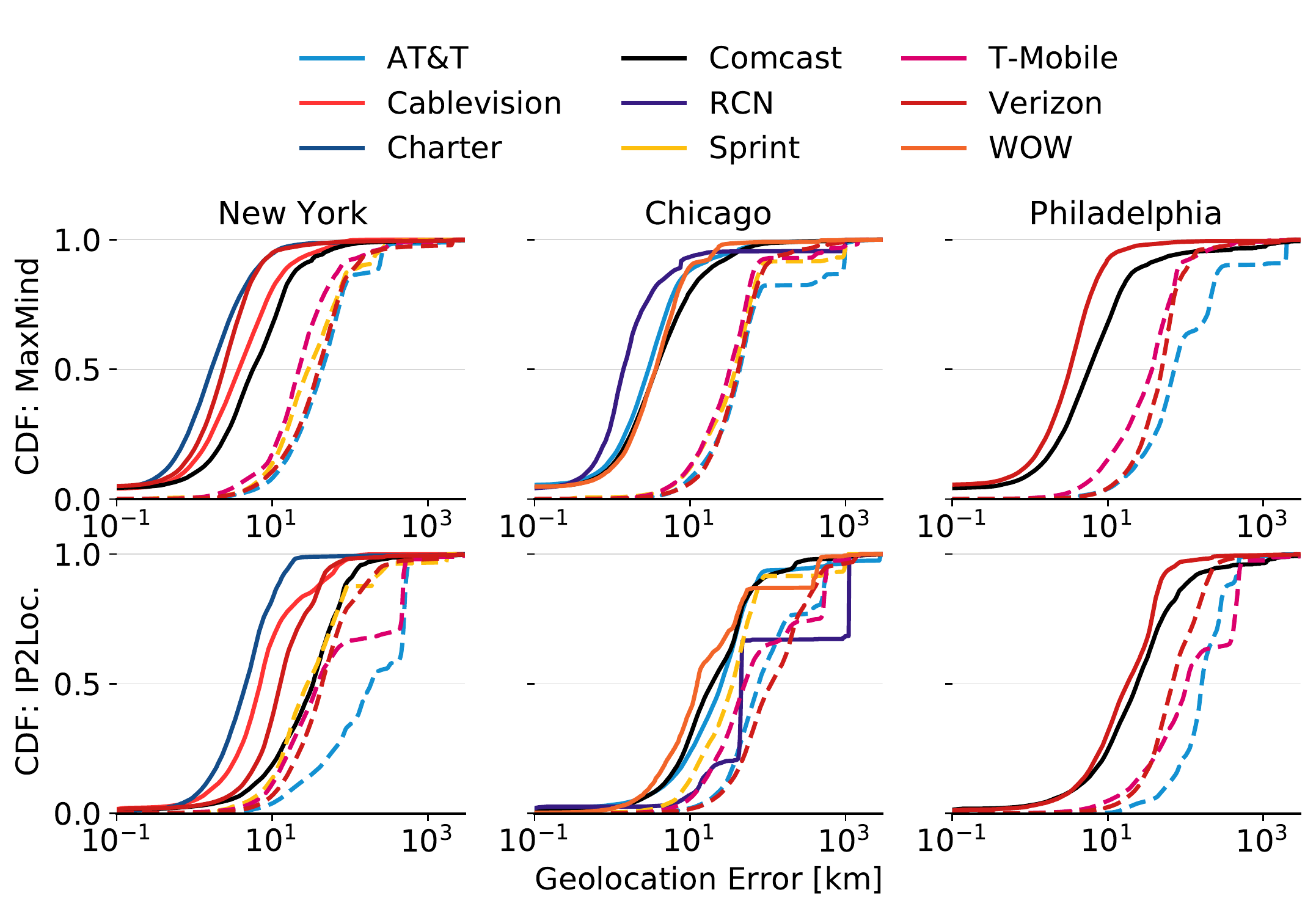}
\caption{Geolocation performance by city, database provider, and ISP. The Figure is identical to Figure~\ref{fig:cdf_cities_dbs} of the text, except that ISPs are identified via whois lookups, rather than by ASN.  The ASNs associated with each ISP are listed in Table~\ref{tab:isp_asn}; note that this list responds to the ASNs seen in the data, and it is not exhaustive for each ISP.  Free versions of the database are used in each case.  ISPs are shown by their ``brand" colors, according to the whois database, which leaves the Sprint and T-Mobile networks distinguishable. Fixed-line networks are denoted by solid lines while mobile networks shown by dashed lines. \label{fig:cdf_cities_asn}}
\end{figure}

\begin{table}[h!]
\centering
\begin{tabular}{cl}
\hline
ISP & ASNs \\
\hline
AT\&T                     & 7018, 2386, 6389, 2686, 4473, 4466, 797, 6431, 17225, 17227 \\
AT\&T Mobile              & 20057 \\
Cablevision               & 6128, 13490, 32953, 14638, 19720 \\
\multirow{2}{*}{Charter}  & 12271, 10796, 20115, 11351, 11426, 33363, 20001, \\
                          & 11427, 33588, 14065, 7843, 17359, 16787 \\
\multirow{2}{*}{Comcast}  & 7922, 33491, 33659, 33287, 7016, 33657, 33651, 7725, \\
                          & 7015, 20214, 33661, 395980, 33652, 396019, 396021 \\
RCN                       & 6079 \\
Sprint                    & 10507, 1239 \\
T-Mobile                  & 21928 \\
Verizon                   & 701, 2828, 23148, 15133, 11486, 12079 \\
Verizon Mobile            & 22394, 6256, 6167 \\
WOW!                      & 12083, 11693 \\
\hline
\end{tabular}
\caption{
  Autonomous systems asssociated with each ISP, 
    for the data within the study region.
  This listing is a categorization of the ASNs seen most-frequently 
    in the data and it is not expected to be an exhaustive listing
    of all ASes corresponding to the ISPs, even in
    the New York, Chicago, and Philadelphia regions.
  ASN ordering is according to the number of subnets associated with each AS, in the data.
  \label{tab:isp_asn}}
\end{table}

\end{sloppypar}
\end{document}